\documentclass[twocolumn,secnumarabic,amssymb, aps, prl]{revtex4-1}
\usepackage{graphicx}
\usepackage{amsmath}
\usepackage{amsfonts}
\usepackage{slashed}
\usepackage{multirow}
\usepackage{appendix}
\usepackage[colorlinks, linkcolor=blue, anchorcolor=blue,citecolor=blue]{hyperref}
\usepackage{float}

\begin{document}
	\title{Critical Behavior and Duality in Dimensionally Reduced Planar Chern-Simons Superconductors}%

\author{Yi-Hui Xing$^{1,2}$}
\author{Lin Zhuang$^{3}$}
\email{stszhl@mail.sysu.edu.cn}
\author{E. C. Marino$^{4}$}
\author{Wu-Ming Liu$^{1,2,5}$}
\email{wliu@iphy.ac.cn}
\affiliation{$^{1}$Beijing National Laboratory for Condensed Matter Physics, Institute of Physics, Chinese Academy of Sciences, Beijing 100190, China}
\affiliation{$^{2}$School of Physical Sciences, University of Chinese Academy of Sciences, Beijing 100190, China}
\affiliation{$^{3}$State Key Laboratory of Optoelectronic Materials and Technologies, School of Physics, Sun Yat-Sen University, Guangzhou 510275, China}
\affiliation{$^{4}$Instituto de Física, Universidade Federal do Rio de Janeiro, Rio de Janeiro RJ 21941-972, Brazil}
\affiliation{$^{5}$Songshan Lake Materials Laboratory, Dongguan, Guangdong 523808, China}

\begin{abstract}
	Tha quantum electrodynamics of particles constrained to move on a plane is not a fully dimensionally reduced theory because the gauge fields through which they interact live in higher dimensions. By constraining the gauge field to the surface of the bulk, we obtain a fully reduced planar Abelian Chern-Simons Higgs model that can describe the vortex dynamics and second-order superconducting-normal phase transitions in planar Chern-Simons superconductors. Dual analyses performed before and after dimensional reduction yield the same Lagrangian for describing the vortex dynamics, indicating the self-consistency of our reduced theory. Compared to ordinary (2+1)-dimensional electrodynamics, we obtain anomalous fermion statistical vortices, consistent with results considering boundary effects. An additional electric charge constraint and different Chern-Simons parameter constraints are also found, which may help define a self-dual conformal field theory. Our renormalization group analysis shows that the quantized critical exponent depends on the Chern-Simons parameter. Quench disorder can bring more stable fixed points with different dynamical critical exponents. If we dimensionally reduce to a curved surface, our theory can also be extended to curved spacetimes, where geometric flow will be introduced and compete with vortex flow.

\end{abstract}

\maketitle

\textit{Introduction.}
The discoveries of the quantum Hall effect \cite{WOS:000233133500043,WOS:000233458500054,PhysRevLett.95.146802,RevModPhys.81.109} and topological planar materials \cite{ WOS:000277210500157,RevModPhys.82,RevModPhys.88.035005,WOS:000748544900007} in the past 20 years have attracted attention to (2+1)-dimensional [(2+1)D] materials.  
Electromagnetic fields play an important role in these planar materials \cite{PhysRevB.78.195424,WOS:000385287800001,WOS:000704967300002, WOS:000862090100001,WOS:000787388900009,PhysRevA.92.043609,Sun_2013,PhysRevLett.130.206701}, prompting the investigation of effective theories that incorporate gauge fields.
The direct application of quantum electrodynamics in (2+1)D (QED3) in order to describe the electromagnetic interaction in such planar materials leads to incorrect results because despite the fact that the quasiparticles are constrained to a plane, the gauge field is not. 
For this reason, the gauge fields in QED3 cannot be considered completely fully fledged dimensionally reduced from (3+1)D quantum electrodynamics. 
Pseudoquantum electrodynamics (PQED), which is also called reduced quantum electrodynamics, is the correct way for introducing the $U(1)$ gauge fields in planar materials, that in spite of being a fully (2+1)D theory, does describe a (3+1)D electromagnetic field interacting with planar particles. 
PQED was first proposed by Marino \cite{MARINO1993551} and has attracted increasing attention both in theoretical \cite{WOS:000488071200096,PhysRevD.87.125002, WOS:000655903100010,WOS:000766932700005,PhysRevD.102.125032} and numerical \cite{WOS:000600849000006} research during these years. 

PQED avoids the logarithmic divergent of a Coulomb potential in the plane which is associated with QED3. In the static limit, the Coulomb potential $V(r)\propto 1/r$ is obtained in PQED and it can correctly describe the electron-electron interactions in two-dimensional materials. In PQED, the normal Maxwell term $F_{\mu\nu}^2$, with gauge field strength $F_{\mu\nu}$, is changed into a nonlocal term $F_{\mu\nu}^2/\sqrt{\Box}$. It has been proved that the PQED is unitary \cite{WOS:000345174400003} and the causality is preserved  \cite{do1992canonical}.

Planar systems contain many interesting phenomena, such as superconductivity \cite{PhysRevB.42.4036,CHEN-1989,liu2023pair},  anyon statistics \cite{WOS:000627412700002,WOS:A1992JE92800038,PhysRevD.44.441,10.1007/3-540-46637-1_3,wilczek1990fractional,PhysRevLett.61.2827}, quantum Hall effect \cite{WOS:000399816500018,WOS:000463902800014}, and quantum vortices \cite{WOS:000406559000011,WOS:000757932400001}. They can all result from the Chern-Simons (CS) term, which can be obtained from the one-loop correction of the gauge field in Fermi quantum electrodynamics. It has no further corrections at higher loops \cite{WOS:A1985ASE0100024}, a result known as the Coleman-Hill theorem. Gauge invariance is restored by using Pauli-Villars regularization while leading to a parity anomaly \cite{WOS:A1984SQ67600020,WOS:A1984RW82500006}. Due to the natural introduction of the CS term through the integration of the electronic field in superconducting systems, and the intrinsic vortex excitations, we choose to study planar CS superconductors.

In this Letter, we demonstrate that the planar Abelian CS Higgs model, which describes planar CS superconductors, exhibits a second-order superconducting-normal phase transition. This planar theory is significantly different from QED3, as the CS term is no longer a topological mass term and exhibits distinct critical behaviors. Considering the inevitable existence of impurities in a real experimental system, we conduct a renormalization group (RG) analysis in the presence of weak quenched disorder, and identify that the fixed points and critical exponents depend on the CS parameter, while disorder can yield more diverse critical behaviors. Finally, based on the aforementioned phase transition behaviors and statistical effects due to the CS term, it is feasible to discuss the properties of vortices. We obtain a unified effective action, which describes vortex loops, based on two different considerations from the surface and bulk, revealing that the anomalous fermionic statistics of vortices are consistent with the inclusion of boundary effects, and our results do not rely on the limit $e^2\rightarrow \infty$.

\begin{figure}
	\centering
	\includegraphics[width=\columnwidth]{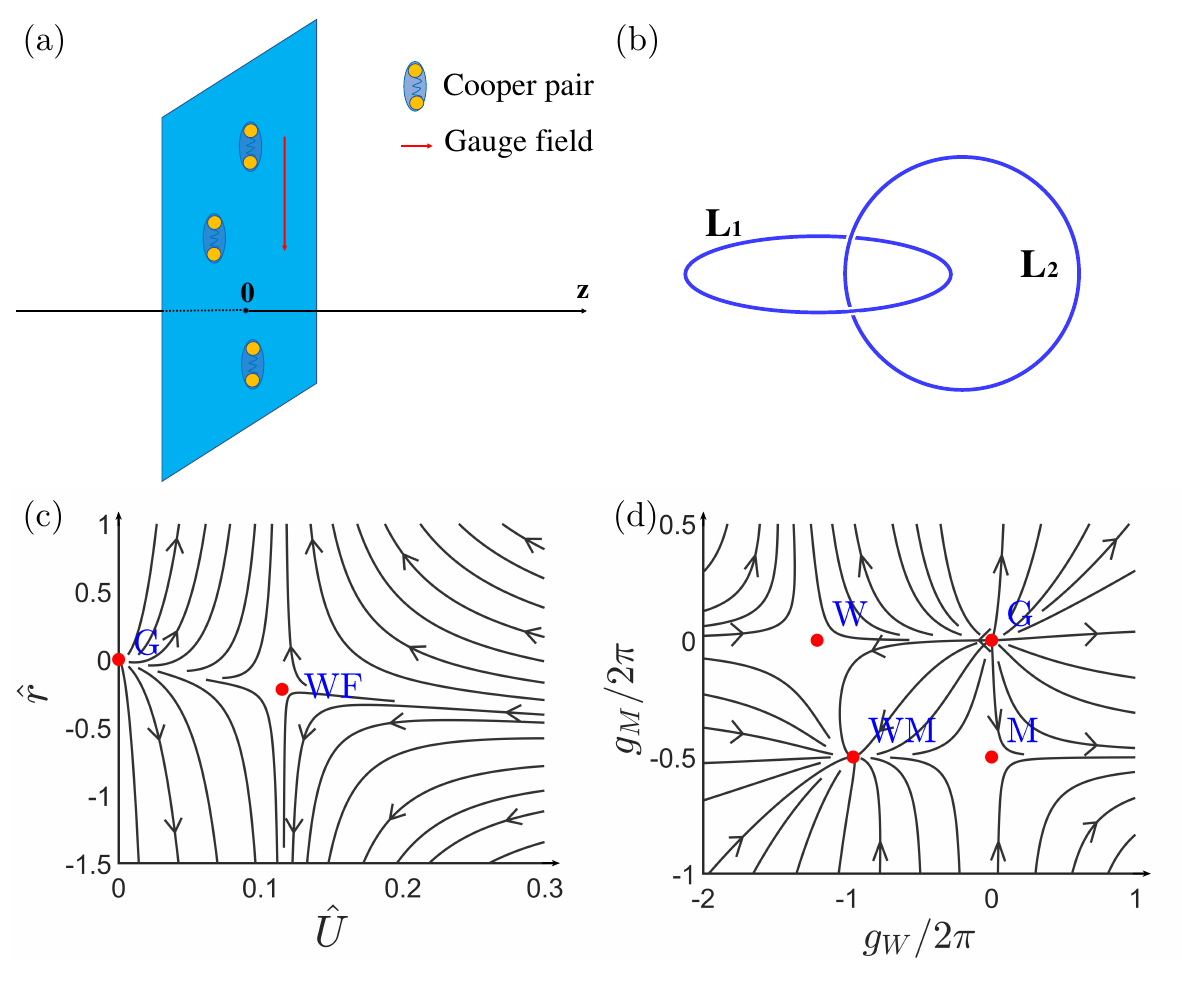}
	\caption{
		(a) Schematic illustration of planar Chern-Simons superconducting system in electromagnetic fields. Cooper pairs are constrained to the plane and the electromagnetic gauge field also lives in (2+1)D. (b) Linking of two vortex worldlines $\mathrm{L_1}$ and $\mathrm{L_2}$. (c) RG flow of the dimensionless couplings $\hat{U}$ and $\hat{r}$ in the absence of disorder, where $n=N=1$. There is a Gaussian fixed point (G) and a Wilson-Fisher fixed point (WF) stable along $\hat{U}$ direction. (d) RG flow in the ($g_W/2\pi$, $g_M/2\pi$) plane with weak quenched mass disorder and current disorder, where $c=25/558$ and all other couplings are set to zero. Besides the Gaussian fixed point (G), the other three fixed points (W, M and WM) are stable along the directions of $g_W$ , $g_M$ and both respectively.
	}
	\label{Fig1}
\end{figure}

\textit{Dimensionally reduced planar Chern-Simons superconductors.} The dimensionally reduced planar Abelian CS Higgs model, which describes the superconducting-normal phase transition in a planar system, reads as
\begin{equation}
	\begin{aligned}
		\mathcal{L}=&\frac{1}{4e^2}F^{\mu\nu}\frac{2}{\sqrt{\Box}}F_{\mu\nu}+|(\partial_\mu -iA_\mu)\phi|^2+r|\phi|^2 \\
		&+\frac{U}{2}|\phi|^4 +\xi A^\mu\frac{\partial_\mu \partial_\nu}{\sqrt{\Box}}A^\nu+i\frac{\theta}{2}\epsilon^{\mu\nu\rho}A_\mu\partial_\nu A_\rho,
	\end{aligned}
	\label{la}
\end{equation}
where $F_{\mu\nu}=\partial_\mu A_\nu-\partial_\nu A_\mu$ with $A_\mu$ being the $U(1)$ gauge field. Note that, since $A_\mu$ lives in (2+1)D and interacts with the massive planar complex scalar field $\phi$, as shown in Fig. \ref{Fig1}(a), but corresponding to a $U(1)$ electromagnetic gauge field living in (3+1)D, we use the nonlocal PQED theory to describe it. $e$ is the effective electric charge with zero scalar dimension ($[e]=0$), $\Box$ is the
d’Alembertian operator, and $\xi$ is a gauge fixing parameter, where $\theta=n/2\pi$ is the CS parameter with zero scaling dimension ($[\theta]=0$) and $n$ must be an integer if the $U(1)$ gauge ﬁeld is compact \cite{compacegaugefield} and the charged excitations all have integer charge and are fully gapped. We note that the CS term is not gauge invariant except for the topologically trivial gauge transformation, while $e^{-S_{\mathrm{CS}}}$ is always gauge invariant with integer CS level $n$.

Our attention is drawn to the model above, which describes the planar superconducting system, motivated by the following considerations: (i) The $|\phi|^4$ theory, $\mathcal{L}_{\phi}=|\partial_\mu\phi|^2+r|\phi|^2+U|\phi|^4/2$, can capture the second-order superconducting-normal phase transition \cite{ginzburg2009theory} without an electromagnetic field. The complex scalar field $\phi$ represents Cooper pairs. If we turn on the electromagnetic field and consider it interacts with Cooper pairs which are constrained on the plane, it can be described by the nonlocal planar Abelian Higgs model. (ii) Considering the massive electrons near the Dirac point on the plane interact with an electromagnetic field, the Lagrangian is given by $\bar{\psi}(\gamma^\mu\partial_\mu-i\gamma^\mu A_\mu-M)\psi$, where $M$ is the mass of the electrons. Using Pauli-Villars regularization and integrating out the Fermi field, we get the CS term \cite{niemi1983axial,WOS:A1984SQ67600020}: $\mathcal{L}_{F}=i\mathrm{sign}(M)\epsilon^{\mu\nu\rho}A_\mu\partial_\nu A_\rho/8\pi$. We only consider the electrons near the Fermi surface and they always appear in pairs to form Cooper pairs, so only even multiples of $\mathcal{L}_{F}$ appear in Lagrangian (\ref{la}) \cite{Supplimentary_material}. In addition, the CS term may also induce the fractional (anyon) statistics which attracts our attention. (iii) The reason why we do not consider the interaction $\lambda |\psi|^4$ is that the scaling dimension of $\lambda$ is smaller than 0 ($[\lambda]=2-D$) and is irrelevant. The terms such as $g\phi \bar{\psi}\bar{\psi}$ will renormalize the mass term of $\phi$ after integrating out the Fermi field in a large momentum cut.

The bare propagator in the momentum space of the gauge field in Lagrangian (\ref{la}) is	
\begin{equation}
	\begin{aligned}
		\Delta^0_{\mu\nu}(p)=&\frac{2e^2}{4+\theta^2e^4}\frac{\delta_{\mu\nu}}{\sqrt{p^2}}-\frac{\theta e^4}{4+\theta^2 e^4}\frac{\epsilon_{\mu\nu\alpha}p^\alpha}{p^2} \\
		&-(\frac{2 e^2}{4+\theta^2 e^4}-\frac{e^2}{\xi})\frac{p_\mu p_\nu}{p^2\sqrt{p^2}}.
	\end{aligned}
\end{equation}
Without the CS term ($\theta=0$), the equation describes the Coulomb interaction in plane \cite{coulombpotential} and it has the same form as the leading-loop renormalization photon propagator at the limit of large-$N$ in Fermi massless QED3 theory \cite{WOS:A1986C803100039,thomson2017quantum}. The CS term no longer acts as the topological mass term of the gauge field as in QED3 theory, instead playing a role in regulating the intensity of the Coulomb interaction.

\textit{Renormalization group analysis without disorder.} We will use the Landau gauge ($\xi=\infty$) for convenience in this part. The one-loop RG $\beta$ equations (details are presented in the Supplemental Material \cite{Supplimentary_material}),
\begin{equation}
	\begin{aligned}
		&\beta_{\hat{r}}\!=\!\mu\frac{d\hat{r}}{d\mu}\!=\!\!-\!(2+\frac{4\hat{e}^2 }{3\pi^2(4+\hat{\theta}^2\hat{e}^4)})\hat{r}-2(N+1)\hat{U},\\
		&\beta_{\hat{U}}\!=\!\mu\frac{d\hat{U}}{d\mu}\!=\!\!-\!(4\!-\!D\!+\!\frac{8\hat{e}^2 }{3\pi^2(4\!+\!\hat{\theta}^2\hat{e}^4)})\hat{U}\!+\!2(N\!+\!4)\hat{U}^2,\\
		&\beta_{{\hat{e}^2}}\!=\!\mu\frac{d\hat{e}^2}{d\mu}\!=\!\!-\!(3-D)\hat{e}^2+(3-D)\frac{N}{32 }\hat{e}^4
		\label{RGgauge}
	\end{aligned}
\end{equation}
are obtained by defining the dimensionless renormalized couplings $r/\mu^2\rightarrow \hat{r}$, $S_DU/\hat{r}^{(4-D)/2}\mu^{4-D}\rightarrow \hat{U}$, electronic charge $e_r^2/\mu^{3-D}\rightarrow \hat{e}^2$, and CS level $\theta \mu^{3-D}\rightarrow \hat{\theta}$, with $S_D=(4-D)\Gamma(2-D/2)/(4\pi)^{D/2}$ and spacetime dimension $D$.

When $D>3$, $\hat{e}_*^2=0$ is the IR stable fix point, and we get the same RG theory as without the gauge field. When $D<3$, $\hat{e}_*^2=32/N$ is the IR stable fix point as the bare coupling  $e^2\rightarrow \infty$. Inserting the ﬁxed point into the first two $\beta$ functions in Eqs. (\ref{RGgauge}), we get the Gaussian fixed point, $(\hat{r}_G,\hat{U}_G)=(0,0)$, and the $O(2N)$ symmetric Wilson-Fisher fixed point, $(\hat{r}_{WF},\hat{U}_{WF})=(-(N+1)(1+\epsilon+\frac{8\hat{e}_*^2 }{3\pi^2(4+\hat{\theta}^2\hat{e}_*^4)})/(N+4)(2+\frac{4\hat{e}_*^2 }{3\pi^2(4+\hat{\theta}^2\hat{e}_*^4)}),(1+\epsilon+\frac{8\hat{e}_*^2}{3\pi^2(4+\hat{\theta}^2\hat{e}_*^4)})/2(N+4))$, where $\epsilon=3-D$ and $N$ is the total categories of the complex scalar field \cite{doi:10.1142/S0217751X94001448,PhysRevB.87.041401,mackenzie1995planar}. It can effectively describe superconductors when $N=1$. The RG flow for the $\beta$ function of the dimensionless coupling $\hat{r}$ and $\hat{U}$ is shown in Fig. \ref{Fig1}(c). It has similar properties to the $\phi^4$ theory, but with a different critical value. The nature of the critical behaviors \cite{watanabe2012new,lima2019fully} is described by the field theory of the Wilson-Fisher fixed point with a controllable CS level $n$. The critical temperature of the phase transition for one-loop at finite temperature is $T_c\sim (r_{\mathrm{WF}}-r)/U$ \cite{Supplimentary_material}.

For level 1 CS term, we can get the critical exponent $\eta\approx 0.41$ and the IR fixed point is $(-0.26,0.14)$ when we consider the superconductor model ($N=1$). At long distances, the critical pair correlation function has the form $\left \langle \phi(x)\phi(0) \right \rangle \sim 1/|x|^{1.41}$.  All calculations are performed in $D=3-\epsilon$ because the PQED and CS term restrict the dimension of spacetime to three, which is different from QED3, where the RG analysis should be conducted in $D=4-\delta$ with $\delta=1$.

\textit{Renormalization group analysis with weak quenched disorder.} Due to the influence of impurities, experimental systems inherently possess both disorders, so we need to take into account the effects of weak quenched disorders. We only consider disorders coupling to a gauge-invariant operator $\mathcal{O} $, $S_{\mathrm{dis}}[\mathcal{O}]=\int d^dxd\tau D(x)\mathcal{O}(x,\tau)$, where we have denoted the Euclidean spacetime coordinates as $r=(x,\tau)$. The quenched random coupling $D(x)$ is time independent, and is a Gaussian random variable, $\overline{D(x)D(x')}=\frac{g_D}{2}\delta^d(x-x')$  with zero average. $g_D$ is the variance of $D(x)$ and controls the strength of the disorder. We use the replica trick to treat disorder questions.

Based on the RG analysis above, it has been determined that the phase transition is primarily controlled by the mass term. Therefore, we can first consider the mass disorder $\phi^*\phi$ while neglecting the mass term for convenience. The global $U(1)$ symmetry has a conserved current $J_\mu=i(\phi\overleftrightarrow{\partial_\mu}\phi^*)\equiv i(\phi\partial_\mu \phi^* -\phi^*\partial_\mu \phi)$ and we can also have flux disorder. The final disorder action takes the form
\begin{equation}
	\begin{aligned}
		S_{dis}\!=\!&\int d^dxd\tau[V(x)|\phi(x,\tau)|^2\!+\!iW(x)\phi(x,\tau)\overleftrightarrow{\partial_0}\phi^*(x,\tau) \\
		&\!+\!iM^i(x)\phi(x,\tau)\overleftrightarrow{\partial_i}\phi^*(x,\tau)].
	\end{aligned}
\end{equation}
Since disorders break the Lorentz invariance, we need to separate space and time. We use the Roman letters $i$, $j$, $k$, etc., indicating the sum is over the spatial coordinates and the Greek letters $\mu$, $\nu$, $\delta$, etc., including time as well, where $V(x)$, $W(x)$, $M^i(x)$ are Gaussian random variables with a vanishing mean, and the variance is $g_V$, $g_W$, $g_M$, respectively. The average of two of them is vanishing.

We obtained the RG flow equations in one-loop order with dimensional regularization and a Feynman gauge (details are presented in the Supplemental Material \cite{Supplimentary_material}),
\begin{equation}
	\begin{aligned}
		&z=1+\frac{1}{4\pi}(2g_W-g_M),	\\
		&\beta_{g_V}=-g_V(2+\epsilon+\frac{16e^2}{3\pi^2(4+\theta^2e^4)}+\frac{g_M+g_W}{2\pi}),\\
		&\beta_{g_W}=-g_W(\epsilon+\frac{54e^2}{15\pi^2(4+\theta^2e^4)}+\frac{g_M+2g_W}{2\pi}),\\
		&\beta_{g_M}=-g_M(\epsilon+\frac{78e^2}{15\pi^2(4+\theta^2e^4)}+\frac{7g_M}{2\pi}),\\
		&\beta_{U}=-U(1+\epsilon-\frac{8e^2}{3\pi^2(4+\theta^2e^4)}-\frac{g_M}{\pi}),\\
		&\beta_{e^2}=-e^2\epsilon(1-\frac{Ne^2}{16}),\\
	\end{aligned}
\end{equation}
where $g_V$, $g_W$, $g_M$, $U$, $e^2$, $\theta$ are all the dimensionless renormalized couplings and $z$ is the dynamical critical exponent. The quenched disorders bring richer critical behaviors \cite{PhysRevB.95.235145,PhysRevLett.65.923}. The RG flow in the $(g_W,g_M)$ plane is shown in Fig. \ref{Fig1}(d) with $c=e^2/15\pi^2(4+\theta^2 e^4)=25/558$, $e^{*2}=16/N$, and all other couplings are set to zero. There are three IR stable fixed points. One of them, $(g_M,g_W,z)=(-156\pi c/7,0,1+39c/7)$, has a dynamical critical exponent larger than 1 and the others, $(g_M,g_W,z)=(0,-54\pi c,1-27c)$ and $(g_M,g_W,z)=(-156\pi c/7,-300\pi c/7,1-111c/7)$, have a dynamical critical exponent smaller than 1. If we consider an effective theory near the Dirac point and view the scalar field as an order parameter \cite{Supplimentary_material}, the damping of quasiparticles $\sim \omega^{1/z}$ at the critical point. Then the change in the dynamic critical exponent will lead to the transition of the scaling from a Fermi liquid to a non-Fermi liquid \cite{PhysRevLett.130.083603,PhysRevB.103.235129,PhysRevB.106.115151,sachdev2023quantum}. This conclusion is also reasonable because, in our calculations involving quenched disorder, we have kept the boson mass at zero ($r=0$). Moreover, this fixed point only emerges when the bosonic potential disorder interaction vanishes, i.e., $g^*{V}=0$, which corresponds to an unstable fixed point of $g_{V}$. All these observations are consistent with the conclusion that the non-Fermi liquid appears at the quantum critical point.

\textit{Vortices and duality.} To discuss the properties of vortices, we change the complex scalar field in Lagrangian (\ref{la}) into polar coordinates $\phi=\rho e^{i\theta}$ with a superfluid density $|\rho|^2$ and phase $\theta$. By a dual analysis in (2+1)D [surface of the (3+1)D bulk] or electromagnetic duality in the bulk (pull back to the surface) as shown in the Supplemental Material \cite{Supplimentary_material}, we can get the effective Lagrangian describing the insulator (vortices condense to give an insulator \cite{PhysRevLett.65.923}),
\begin{equation}
	\begin{aligned}	
		\mathcal{L}_{eff}=&\frac{1}{16\pi^2|\rho|^2}(\epsilon^{\mu\nu\lambda}\partial_\mu a_\lambda)^2+\frac{e^2}{4\pi^2(4+\theta^2e^4)}\frac{(\epsilon^{\mu\nu\lambda}\partial_\mu a_\lambda)^2}{\sqrt{\Box}}\\
		&-i a^\lambda J_\lambda^a-i\frac{\theta e^4}{8\pi^2(4+\theta^2e^4)}\epsilon^{\mu\nu\lambda}a_\mu\partial_\nu a_\lambda,
		\label{eff}
	\end{aligned}
\end{equation}
where $a_\mu$ is the dual $U(1)$ gauge ﬁeld and the $J_\mu^a$ is the vortex current. The vortices are minimally coupled to $a_\mu$. The fact that these two methods yield the same results demonstrates the validity of our reduced nonlocal theory. One notable difference from QED3 is that our effective vortex Lagrangian contains both a normal Maxwell term and a nonlocal one [the first and second terms in effective Lagrangian (\ref{eff}), respectively].

In real space the vortex current can be written as \cite{shyta2021deconfined,kleinert1989gauge,turker2020bosonization}
\begin{equation}
	J_\mu^a(x)=\epsilon_{\mu\nu\lambda}\partial^\nu v^\lambda(x)=\sum_cn_c\oint_{L_c}dy^{c}_\mu\delta^3(x-y^{c}),
	\label{j}
\end{equation}
with a vortex charge $n_c\in\mathbb{Z}$ and the vortex loop $L_c$. Under the limit of $|\rho|^2\gg 1 $ or $p^2\rightarrow 0$, we can ignore the normal Maxwell term and integrate over the dynamical gauge fields $a_\mu$. This yields the effective action for the vortex current $J_\mu^a$, i.e.,
\begin{equation}
	\begin{aligned}	
		S_{eff}=&\int d^3x d^3x'
		[-\frac{4\pi^2}{e^2}\frac{J_\mu^a(x) J^{\mu a}(x')}{2\pi^2|x-x'|^2}\\
		&+i2\pi^2 \theta\frac{\epsilon^{ \mu\alpha\nu} J_\mu^a(x)(x-x')_\alpha 		J_\nu^a(x')}{4\pi|x-x'|^3}].
		\label{eact}
	\end{aligned}
\end{equation}

Substituting Eq. (\ref{j}) into action (\ref{eact}), we find that the second term on the right-hand side can characterize $2\pi^2\theta$ times the linking number \cite{shyta2021deconfined,turker2020bosonization} of two different vortex loops as shown in Fig. \ref{Fig1}(b). Such an effective self-statistical angle of the unit vortex is $2\pi^2  \theta$ and equal to $\pi n$, satisfying the Fermi statistics at level 1. It is just the statistical anomaly after considering the bulk effect as analyzed in Refs. \cite{metlitski2013bosonic,ye2013symmetry}. In QED3, this result is obtained under the limit of $e^2\rightarrow\infty$ \cite{shyta2021deconfined,metlitski2013bosonic}, but ours does not rely on this limit.

\begin{table}[b]
	\centering	
	\caption{Table showing the duality between Lagrangian (\ref{la}) and (\ref{dua}). The central row gives the constraints of the couplings $e^2$ and $\theta$. The last row gives the phases of a Lagrangian dual to the other.}.	
	\begin{ruledtabular}		
		\begin{tabular}{ccc}			
			& $\mathcal{L}[A_\mu,\phi]$ &  $\tilde{\mathcal{L}} [{\tilde{A}}_\mu,\tilde{\phi}] $\\			
			\colrule			
			\multirow{2}{*}{Couplings} & $e^2$ & ${\tilde{e}}^2=\pi^2\left(4+\theta^2e^4\right)/e^2$ \\
			&$\theta$ & $\tilde{\theta}=-\theta e^4/4\pi^2\left(4+\theta^2e^4\right)$ \\
			
			\multirow{2}{*}{Phases} & Superconductor & Insulator \\
			&Normal  & Superfluid \\
			
		\end{tabular}		
	\end{ruledtabular}	
	\label{tab}
\end{table}

The conservation of current $J_\mu^a$, $\partial^\mu J_\mu^a=0$, indicates we can add the term $i\partial^\mu J_\mu^a \xi$ as a constraint setting. At the same time, introducing a convergence factor \cite{peskin1978mandelstam} $tJ_\mu^2/2$ can be viewed as the chemical potential of the vortex loops \cite{shyta2022frozen}. We can get the same formal structure as the original Lagrangian (\ref{la}) at long wavelengths after integrating over $J_\mu$:
\begin{equation}
	\begin{aligned}
		\mathcal{\tilde{L}}_{dual}=&\frac{1}{4\tilde{e}^2}\tilde{F}^{\mu\nu}\frac{2}{\sqrt{\Box}}\tilde{F}_{\mu\nu}+|(\partial_\mu -i\tilde{A}_\mu)\tilde{\phi}|^2+\tilde{r}|\tilde{\phi}|^2 \\
		&+\frac{\tilde{U}}{2}|\tilde{\phi}|^4+i\frac{\tilde{\theta}}{2}\epsilon^{\mu\nu\rho}\tilde{A}_\mu\partial_\nu \tilde{A}_\rho.
	\end{aligned}
	\label{dua}
\end{equation}
A significant difference from the QED3 is that the nonlocal kinetic energy of the gauge field in the dual Lagrangian (\ref{dua}) is independent of the kinetic energy of the bosonic field $\phi$ in the original Lagrangian (\ref{la}). In the long-wavelength limit, the related normal Maxwell term is much smaller than the nonlocal one and is discarded. The duality transformations show it satisfies $\tilde{\theta}=-\theta e^4/[4\pi^2 (4+\theta^2e^4)]$, which is different from the QED3-Abelian Higgs model \cite{burgess2001particle,herzog2007quantum}, $\tilde{\theta}\theta=-1/4\pi^2$. However, in the limit of $e^2\rightarrow \infty$, the two results match. In addition, we have another constraint, $\tilde{e}^2e^2=\pi^2(4+\theta^2e^4)$. It is possible \cite{Higgsmodel} to set both $e^2=\infty$ and $\tilde{e}^2=\infty$ because the kinetic terms of the dual gauge fields $\tilde{A}_\mu$ are not dual to the kinetic energy of the $\phi$ particle compared with QED3, reflecting the self-duality of the planar Abelian CS Higgs model. Then the remaining CS term can induce a $2\pi/n$ flux bound to each $\phi$ bosonic particle worldline and $-2\pi n$ for dual bosonic field $\tilde{\phi}$. It is worth noting that these dualities are only valid in the infrared. As listed in Table \ref{tab}, the superconducting phase of Lagrangian (\ref{la}) is dual to the insulator phase  (with vortex condensation) of Lagrangian (\ref{dua}), while the normal phase is dual to the superfluid phase (with vortex excitation) \cite{Supplimentary_material}. 

It is remarkable that we can also get $O(2N)$ and $O(2N-2k)\times O(2k)$, $0<k<N$, symmetric fixed points \cite{shyta2021deconfined,plischke1994equilibrium,criticle} if we change the interaction from $U\sum_i|\phi_i|^4$ to $U_1\sum_{i<k+1}|\phi_i|^4+2U_2\sum_{i<k+1,j>k}|\phi_i|^2|\phi_j|^2+U_3\sum_{i>k}|\phi_i|^4$, where different values of $N$ can correspond to different theories. It can develop a lot of interesting phenomena, such as the deconfined quantum critical point \cite{ WOS:000220000100033}, and the duality between the QED3–Gross-Neveu theory \cite{WOS:000414668300002}. On the other hand, different scalar fields can be seen as distinct order parameter fields, and their competition can lead to diverse vortex dynamics \cite{PhysRevLett.130.226002}.

In (3+1)D, $F_{\mu\nu}$ can be interpreted as the primary field in free Maxwell theory and it is conformally invariant \cite{ELSHOWK2011578}. However, this is not the case in other dimensions and we can replace it by a conformal gauge action $F_{\mu\nu}^2/\sqrt{\Box}$ in dimension three \cite{WOS:000371250700011,primary}. It is just the result after considering the one-loop fermion vacuum polarization diagram in QED3 theory as mentioned above. Marino \cite{MARINO1993551} looked at it from another perspective, where particles conserved current $j^\mu$ equal to zero when $\mu=3$ and $j^\mu \delta(x^3)$ otherwise, and obtained the same result. We can simply generalize this result to $j^\mu \delta(f(x^i))$, where particles constrain at any (2+1)-dimensional closed (or one-point compactifiable) submanifold $(t,x^i)$ with $f(x^i)=0$. At this time, covariant derivatives with a geometric connection will appear in the Lagrangian and can induce a geometric current in the effective vortex action \cite{PhysRevLett.129.016801,PhysRevLett.124.197001}.  


\textit{Conclusion.} We have studied an Abelian Chern-Simons Higgs model that undergoes a dimensional reduction to the plane, and obtained its quantized critical behaviors. Disorder can be introduced to adjust the critical behavior of the interaction from relevant to irrelevant. If dimensionally reduced to a curved surface, we may obtain a curved PQED model to investigate the critical behaviors of the theory as well as the dynamic behaviors of vortices. Furthermore, self-duality imposes constraints on the charge and CS parameter, which also apply to the flow-flow correlation functions near the critical points \cite{Supplimentary_material,burgess2001particle,PhysRevB.54.4953,witten2005sl}. These constraints can even be extended to nonzero temperature cases. In summary, our reduced theory provides a good idea for the further study of self-duality near the quantum critical point , and can even be extended to nonzero temperature and nonflat spacetime.

\begin{acknowledgments}
	This work was supported by National Key R\&D Program of China under Grants No. 2021YFA1400900, No. 2021YFA0718300, and No. 2021YFA1402100, NSFC under Grants No. 12174461, No. 12234012, No. 12334012, and No. 52327808, Space Application System of China Manned Space Program. E.C.M. has been partially supported by CNPq, FAPERJ, and CAPES.
\end{acknowledgments}
		
	\bibliography{paper.bib}

\begin{thebibliography}{84}
\bibitem [{com()}]{compacegaugefield}%
  \BibitemOpen
  \href@noop {} {}\bibinfo {note} {If the system can be realized in a (2+1)D lattice model, the $U(1)$ gauge field can be compact. In the condensed matter, a compact abelian gauge field means a monopole operator is included in the action.}\BibitemShut {Stop}%
\bibitem [{cou()}]{coulombpotential}%
  \BibitemOpen
  \href@noop {} {}\bibinfo {note} {The coulomb potential or static potential is generated by $D_{00}(p_0=0, \vec{p})$, namely: $V(r)=\frac{e^2}{2}\int \frac{d^2p}{(2\pi)^2}\frac{e^{i\vec{p}\vec{r}}}{|\vec{p}|}=\frac{e^2}{4\pi r}$.}\BibitemShut {Stop}%
\bibitem [{pri()}]{primary}%
  \BibitemOpen
  \href@noop {} {}\bibinfo {note} {However, the gauge field strength $F_{\mu\nu}$ is still not primary. We can assume that it is a descendent of $B$, such as $F_{\mu\nu}=\epsilon_{\mu\nu\gamma}\partial^\rho B$ at level 1.}\BibitemShut {Stop}%
\end{thebibliography}

\clearpage


\makeatletter
\renewcommand{\theequation}{S\arabic{equation}}
\setcounter{equation}{0}
\renewcommand{\thefigure}{S\arabic{figure}}
\setcounter{figure}{0}
\renewcommand{\thesection}{S\arabic{section}}

\onecolumngrid

\section{Detailed derivation of the Chern-Simons term in our model}	
We consider the following Lagrangian: 
\begin{equation}
	\mathcal{L}_F={\bar{\psi}}_i\left(\gamma^\mu\partial_\mu-i\gamma^\mu A_\mu-M\right)\psi_i,
\end{equation}
where $\psi_i$ is the Dirac field describing the electrons near the Fermi surface and the $A_\mu$ is the $U(1)$ gauge field. The mass term breaks the parity symmetry. $i=1,\ldots,2N$ is the flavors of the Dirac fermions. The total flavor of even numbers is because the electrons are appeal in pairs to form Cooper pairs. For example, $N=1$ represents the spin up and down of the electrons and Cooper pairs is composed of a pair of electrons with opposite spin. Integrate out the Fermi field and obtained:
\begin{equation}
	S_{eff}\left(M\right)=-2N\ln{\det{\left(\gamma^\mu\left(\partial_\mu-iA_\mu\right)-M\right)}}.
\end{equation}
Use the identity $\text{Tr} \ln A=\ln \det A$. The linear term of $U(1)$ gauge field $A_\mu$ is traceless. Then the lowest order of $A_\mu$ in the remaining terms is:
\begin{equation}
	S_{eff}(M)=-N\int \frac{d^3p}{(2\pi)^3}A^\mu A^\nu \int \frac{d^3q}{(2\pi)^3}\text{Tr}[\frac{1}{\slashed{q}-M}\gamma_\mu\frac{1}{\slashed{p}+\slashed{q}-M}\gamma_\nu].
\end{equation}
Using the method of Feynman parameter, we have:
\begin{equation}
	S_{eff}(M)=-N\int \frac{d^3p}{(2\pi)^3}A^\mu A^\nu \int_0^1 dx \int \frac{d^3q}{(2\pi)^3}\text{Tr}\frac{iM\slashed{p}\gamma_\mu \gamma_\nu+M^2\gamma_\mu \gamma_\nu-\slashed{q}\gamma_\mu\slashed{q} \gamma_\nu +x(1-x)\slashed{p}\gamma_\mu\slashed{p} \gamma_\nu}{(q^2+M^2+x(1-x)p^2)^2}.
\end{equation}
Using the identity $\text{Tr}(\gamma_\mu \gamma_\nu \gamma_\lambda)=2i\epsilon_{\mu\nu\lambda}$, $\text{Tr}(\gamma_\alpha \gamma_\mu \gamma_\alpha \gamma_\nu)=-2\delta_{\mu\nu}$, $\text{Tr}(\gamma_\alpha\gamma_\mu\gamma_\beta\gamma_\nu)=2\delta_{\alpha\mu}\delta_{\beta \nu}+2\delta_{\alpha\nu}\delta_{\beta \mu}-2\delta_{\alpha\beta}\delta_{\mu\nu}$ (The basis of the Euclidean Cliﬀord algebra $C(V^3_{(3)})$ is $\gamma_0=\gamma_5=-i\gamma_1\gamma_2$, $\gamma_1$, $\gamma_2$.), and replace $q_\alpha q_\beta\rightarrow q^2\delta_{\alpha \beta}$, we get:
\begin{equation}
	S_{eff}(M)=-2N\int \frac{d^3p}{(2\pi)^3}A^\mu A^\nu \int_0^1 dx \int \frac{d^3q}{(2\pi)^3}\frac{x(1-x)(2p_\mu p_\nu-p^2\delta_{\mu\nu})+M^2\delta_{\mu\nu}-M\epsilon_{\mu\nu\lambda}p^\lambda+q^2\delta_{\mu\nu}}{(q^2+M^2+x(1-x)p^2)^2}.
\end{equation} 
On the momentum integral after Taylor expansion of $x$, and select UV truncation as $\Lambda$. We obtained:
\begin{equation}
	S_{eff}(M)=-2N\int \frac{d^3p}{(2\pi)^3}A^\mu A^\nu (-\frac{\text{sgn}(M)}{8\pi}\epsilon_{\mu\nu\lambda}p^\lambda+\frac{\Lambda \delta_{\mu\nu}}{2\pi^2}+\frac{p_\mu p_\nu-p^2\delta_{\mu\nu}}{24\pi\sqrt{M^2}}).
\end{equation} 
After Pauli-Villars normalization and let $M\rightarrow \infty$, we obtained the Chern-Simons term in Eq. (1):
\begin{equation}
	\mathcal{L}_{eff}=\frac{iN}{4\pi}\epsilon_{\mu\nu\lambda}A^\mu\partial^{\lambda}A^\nu.
\end{equation} 
It can bring anyon statistical effects as shown in our letter.

So if we start from an effective theory near a generic Dirac point:
\begin{equation}
	\mathcal{L} _e=\bar{\psi}(\gamma ^{\mu}\partial _{\mu}-i\gamma ^{\mu}A_{\mu}-M)\psi +g\bar{\psi}\bar{\psi}\phi +g^*\psi \psi \phi ^{\ast}+\frac{1}{4e^2}F^{\mu \nu}\frac{2}{\sqrt{\Box}}F_{\mu \nu}+|(\partial _{\mu}\,\,-iA_{\mu})\phi |^2+r|\phi |^2+\frac{U}{2}|\phi |^4.
\end{equation}
We are considering superconducting order $\phi$, and if we take into account charge/spin density wave order, we need to replace the Bose-Fermi coupling term with a Yukawa coupling $\bar{\psi}\psi\phi$. Integrating out all the electrons fields, we obtain the renormalized Lagrangian (1) as shown in the main text.

\section{Renormalization group analysis without disorder  }
\subsection{Zero temperature}
Both the CS term and PQED restrict the space-time dimension to three, so we do the following caculations in $D=3-\epsilon$. We will choose Laudau gauge for convinent .

\begin{figure}[h]
	\centering
	\includegraphics{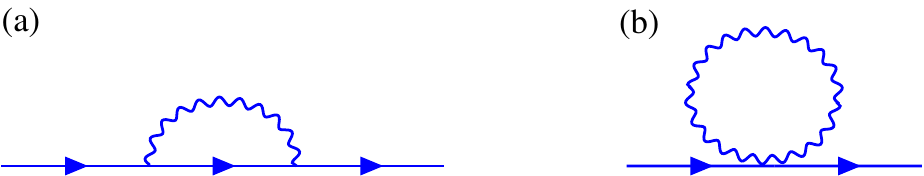}
	\caption{One-loop self-energy diagrams of a scalar field, where we use solid lines for scalar field propagator and sinusoidal lines for optical field propagator.} \label{self-energy}
\end{figure}

To perform RG analysis without disorder, we first need to evaluate the diagram shown in Fig.\ref{self-energy}(a).
\begin{equation}
	\begin{aligned}
		\Sigma_1(p)&=-\int\frac{d^Dk}{(2\pi)^D}\frac{(2p-k)^\mu(2p-k)^\nu}{(p-k)^2+r}\Delta^0_{\mu\nu}(k)
		=-\frac{e^2\mu^\epsilon}{4+\theta^2e^4}\int_0^1\frac{dx}{\sqrt{1-x}}\int\frac{d^dk}{(2\pi)^d}\frac{k^2+p^2(2-x)^2}{(k^2+\Delta_1)^\frac{3}{2}}\\
		&=- \frac{e^2}{(4+\theta^2 e^4)4\pi^2}[\frac{16p^2}{3\epsilon}+{\rm finite}].
	\end{aligned}
\end{equation}	
The second diagram of self energy in Fig.\ref{self-energy}(b) is $0$. After using the minimal subtraction scheme, the wave function renormalization is:
\begin{equation}
	Z_\phi=1+\frac{4e^2}{3(4+\theta^2 e^4)\pi^2}.
\end{equation} 

\begin{figure}[h]
	\centering
	\includegraphics{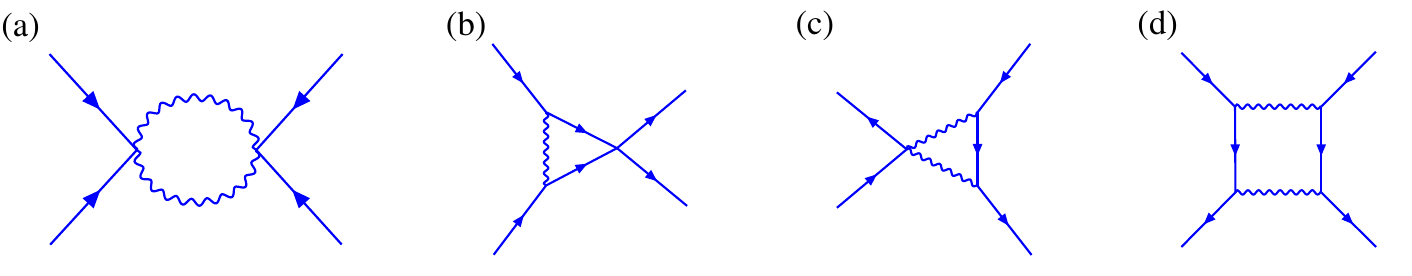}		\ \ \ 	
	\caption{One-loop diagrams contribute to the coupling $U$, where we use solid lines for scalar field propagator and sinusoidal lines for optical field propagator.} \label{coupling}		
\end{figure}
The one-loop diagrams which contribute to the coupling constant $U$ are shown in Fig.\ref{coupling}. The contribution of the first diagram is:
\begin{equation}
	\begin{aligned}
		I(p)=&8\int \frac{d^Dk}{(2\pi)^D}\Delta^{0 \mu\nu}(k)\Delta_{\mu\nu}^0(p-k)\\
		=&\frac{32e^4}{(4+\theta^2 e^4)^2}[\int_0^1\frac{dx}{\sqrt{x(1-x)}} \int \frac{d^Dk}{(2\pi)^D}\frac{1}{k^2+\Delta_{2}}+\int_0^1dx\sqrt{x(1-x)}\int \frac{d^Dk}{(2\pi)^D}\frac{k^4+C(p^2)k^2+x^2(x-1)^2p^4}{(k^2+\Delta_{2})^3}]\\
		&+\frac{32\theta^2 e^8}{(4+\theta^2e^4)^2}\int_0^1dx\int\frac{d^Dk}{(2\pi)^D}\frac{p^2x(x-1)+k^2}{[k^2+\Delta_{2}]^2}\\
		=& -\frac{e^4}{(4+\theta^2 e^4)^2}\sqrt{p^2}(2\theta^2 e^4+\frac{24}{\pi^2}+\frac{3}{\pi}),
	\end{aligned}
\end{equation}	
where $\Delta_{2}=p^2x(1-x)$, $C(p^2)/p^2=(2-x)^2/d+2x(x-1)$. The other one-loop diagrams in Fig.\ref{coupling} vanish when external momentum tends to zero in Landau gauge ($p_\mu(\delta_{\mu\nu}-p_\mu p_\nu)=0$) and have no contribution to $\beta_U$.

\begin{figure}[H]
	\centering
	\includegraphics{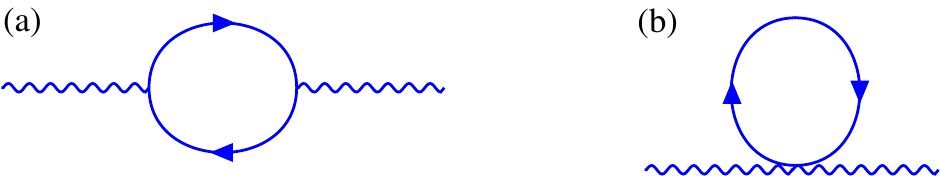}
	\caption{Vacuum polarization diagrams at one-loop, where we use solid lines for scalar field propagator and sinusoidal lines for optical field propagator.} \label{gauge}
\end{figure}
At one-loop order, we have two vacuum polarization diagrams as shown in Fig.\ref{gauge}, the first one reads as:
\begin{equation}
	\begin{aligned}
		\Pi_{\mu\nu 1}(p)&=N\int\frac{d^Dk}{(2\pi)^D}\frac{(p+2k)_\mu(p+2k)_\nu}{(k^2+r)[(p+k)^2+r]}=N\int_0^1dx\int\frac{d^Dk}{(2\pi)^D}[\frac{p_\mu p_\nu (2x-1)^2}{(k^2+\Delta_3)^2}+ \frac{\delta_{\mu\nu}}{d}\frac{4k^2}{(k^2+\Delta_3)^2}]\\
		&=-\frac{ N}{8\pi}[2\sqrt{r}(\delta_{\mu\nu}+\frac{p_\mu p_\nu}{p^2})+\frac{p^2+4r}{\sqrt{p^2}}\arctan{\frac{\sqrt{p^2}}{2\sqrt{r}}}(\delta_{\mu\nu}-\frac{p_\mu p_\nu}{p^2})].\\
	\end{aligned}
\end{equation}

The second one is shown in Fig. \ref{gauge}(b):
\begin{equation}
	\begin{aligned}
		\Pi_{\mu\nu 2}(p)&=-2N\delta_{\mu\nu}\int\frac{d^Dk}{(2\pi)^D}\frac{1}{k^2+r}=\frac{4N}{8\pi}\delta_{\mu\nu}\sqrt{r}.
	\end{aligned}
\end{equation}
Therefore, the one-loop correction to the correlation function of the gauge field is:
\begin{equation}
	\Pi_{\mu\nu}(p)=\Pi_{\mu\nu1}(p)+\Pi_{\mu\nu2}(p)=\left\{\begin{matrix}
		-\frac{Np^2}{24\pi\sqrt{r}}(\delta_{\mu\nu}-\frac{p_\mu p_\nu}{p^2})	& p^2\ll r \\
		-\frac{N\sqrt{p^2}}{16}(\delta_{\mu\nu}-\frac{p_\mu p_\nu}{p^2})	& p^2\gg r.
	\end{matrix}\right.
\end{equation}
We consider the limit $p^2\gg r$, which allows us to avoid the need for analytical extension into the region where $r<0$. The full propagator of the gauge field reads as:
\begin{equation}
	\begin{aligned}
		\Delta_{\mu\nu}(p)=({\Delta^0_{\mu\nu}}(p)^{-1}-\Pi_{\mu\nu}(p))^{-1}=\frac{(2+\frac{N}{16})e^2}{\theta^2e^4+(2+\frac{N}{16})^2}\frac{\delta_{\mu\nu}-p_\mu p_\nu/p^2}{\sqrt{p^2}}-\frac{\theta e^4}{\theta^2e^4+(2+\frac{N}{16})^2}\frac{\epsilon_{\mu\nu\alpha}p^\alpha}{p^2}.
	\end{aligned}
\end{equation}

The effective Pseudo-Maxwell Lagrangian is: $\mathcal{L}_M=(1+\frac{Ne^2}{32})\frac{F_{\mu\nu}F^{\mu\nu}}{2e^2\sqrt{\Box}}$. So the dimensionless renormalized gauge coupling is $\hat{e}^2=e^2\mu^{-\epsilon}/(1+Ne^2\mu^{-\epsilon}/32)$ and reaching the IR fixed point $\hat{e}^2_*=32/N$ when $e^2\rightarrow \infty$. 		

\subsection{Non zero temperature}
According to the Mermin-Wagner-Hohenberg theorem, there are no finite-temperature phase transitions of continuous symmetry breaking in (2+1)D, but they can occur at one-loop in the perturbative expansion using Feynman diagrams. We will using the method in Ref. \cite{WOS:000079575900024} to calculate:

The renormalized coupling can be written as:
\begin{equation}
	R=r+U(N+1)\frac{1}{\beta}\sum_{k_0\ne 0}{\int{\frac{d^2k}{(2\pi )^2}}}\frac{1}{k^2+r}.
\end{equation}
The sum of Matsubara frequency is given as:
\begin{equation}
	R=r+TU(N+1)\int{\frac{xdx}{2\pi}}[\frac{1}{\sqrt{x^2+y}}\frac{1}{e^{\sqrt{x^2+y}}-1}-\frac{1}{x^2+y}+\frac{1}{2\sqrt{x^2+y}}],
	\label{nonzt}
\end{equation}
Where $x=k/T$  and $y=r/T^2$ are dimensionless variables. Note that we did not consider the contribution of self-energy diagram in Fig. \ref{self-energy}, because we can regard $r$  above as renormalized after considering the self-energy diagram in Fig. \ref{self-energy}.

In the low temperature limit $T^2\ll r$ , we need to use the minimum subtraction renormalization scheme to avoid UV divergence, and replace the integral of  $x$ with $[\sqrt{y},\infty]$  to avoid IR divergence. We obtained:
\begin{equation}
	R=r_R-\sqrt{r_R}U(N+1)e^{-\sqrt{r_R}/T}/2\pi ,
\end{equation}	
which is only valid when $r_R\equiv r-r_{WF}>0$  (The first and fourth terms on the right of Eq. (\ref{nonzt}) equal sign form the expression of renormalization of $r$ at zero temperature, and the two can be replaced by $r_R$ .). It captures the gap of the quasi-particles.

In the high temperature limit $T^2\gg r$ , we also use the minimum subtraction renormalization scheme to avoid UV divergence and obtained:
\begin{equation}
	R=r_R+TU(N+1)\zeta (1)/2\pi ,
\end{equation}
where $\zeta(1)$  is the Reimann zeta function. We replace the renormalized parameter $R$ with the correction length $\xi^{-2}$ , and the critical phase transition point is arrived when $\xi=0$ , namely, $T_c\sim -r_R/U$ .

\section{Renormalization group analysis with quenched disorder}
The disorder action under consideration takes the following form:
\begin{equation}
	\begin{aligned}
		S_{dis}=&\int d^dxd\tau[V(x)|\phi(x,\tau)|^2+iW(x)\phi(x,\tau)\overleftrightarrow{\partial_0}\phi^*(x,\tau) +iM^i(x)\phi(x,\tau)\overleftrightarrow{\partial_i}\phi^*(x,\tau)].
	\end{aligned}
\end{equation}
Averaging over disorder, we have
\begin{equation}
	\begin{matrix}
		\overline{V(x)}=0,	&\overline{V(x)V(x')}=\frac{g_V}{2}\delta^d(x-x'), \\
		\overline{W(x)}=0,	&\ \ \overline{W(x)W(x')}=\frac{g_W}{2}\delta^d(x-x'), \\
		\overline{M_i(x)}=0,	&\ \ \ \ \ \ \ \overline{M_i(x)M_j(x')}=\frac{g_M}{2}\delta_{i,j}\delta^d(x-x') ,
	\end{matrix}
\end{equation}
with all other two-points vanishing.

We employ the replica trick and get the disorder-averaged action:
\begin{equation}
	\begin{aligned}
		S_n=&\sum_i \int d^dxd\tau[|(\partial_0 -iA_{i0})\phi_i|^2+|(\partial_k -iA_{ik})\phi_i|^2 +\frac{U}{2}|\phi_i|^4+\frac{1}{4e^2}F_i^{\mu\nu}\frac{2}{\sqrt{\Box}}F_{i\mu\nu}+i\frac{\theta}{2}\epsilon^{\mu\nu\rho}A_{i\mu}\partial_\nu A_{i\rho} \\
		&-\sum_{i,j} \int d^dxd\tau d\tau'[\frac{g_V}{2}\mu^\epsilon |\phi_i(x,\tau)|^2|\phi_j(x,\tau')|^2+\frac{g_W}{2}\mu^\epsilon \phi_i\overleftrightarrow{\partial_0}\phi_i^*(x,\tau)\phi_j\overleftrightarrow{\partial_0}\phi_j^*(x,\tau')\\
		&+\frac{g_M}{2}\mu^\epsilon \phi_i\overleftrightarrow{\partial_k}\phi_i^*(x,\tau)\phi_j\overleftrightarrow{\partial^k}\phi_j^*(x,\tau') ],
	\end{aligned}
\end{equation}
where $i,j$ are the indicators of the replica field and the range of values can be an integer from $1$ to $n$. $n$ is taken to zero at the last of the caculation, so we need only consider the digram which do not vanish in replica field limit. It is remarkable that even though relativistic invariance is explicitly violated, there is no necessity to monitor the relative motion of boson and photon velocities \cite{thomson2017quantum}. This is due to the fact that the low-energy characteristics of the photon propagator solely originate from its interaction with the bosons. In order to employ the dimensional regularization scheme with $D=3-\epsilon$, it is necessary to shift $g_V(g_W,g_M)$ to $g_V\mu^\epsilon(g_W\mu^\epsilon,g_M\mu^\epsilon)$, thereby ensuring that the dimension of the renormalized couplings is $[g_V]=2$, and $[g_W]=[g_M]=0$. 

We define the bare ﬁelds (coordinates) and the renormalized fields (coordinates) satisfied:	
\begin{equation}
	\begin{aligned}
		\begin{matrix}
			\phi_B=Z_2^{1/2}\phi,	& A_{0,B}=Z_{r,0}^{1/2}A_0, &A_{i,B}=Z_{r,xy}^{1/2}A_i,  &\tau_B=\sqrt{Z_2/Z_1}\tau,  &x_B=x
		\end{matrix}
	\end{aligned}
\end{equation}
Gauge invariance gives two constrain $Z{r,xy}=1$ and $Z_2Z_{r,0}=Z_1$.

\begin{figure}
	\includegraphics[width=\columnwidth]{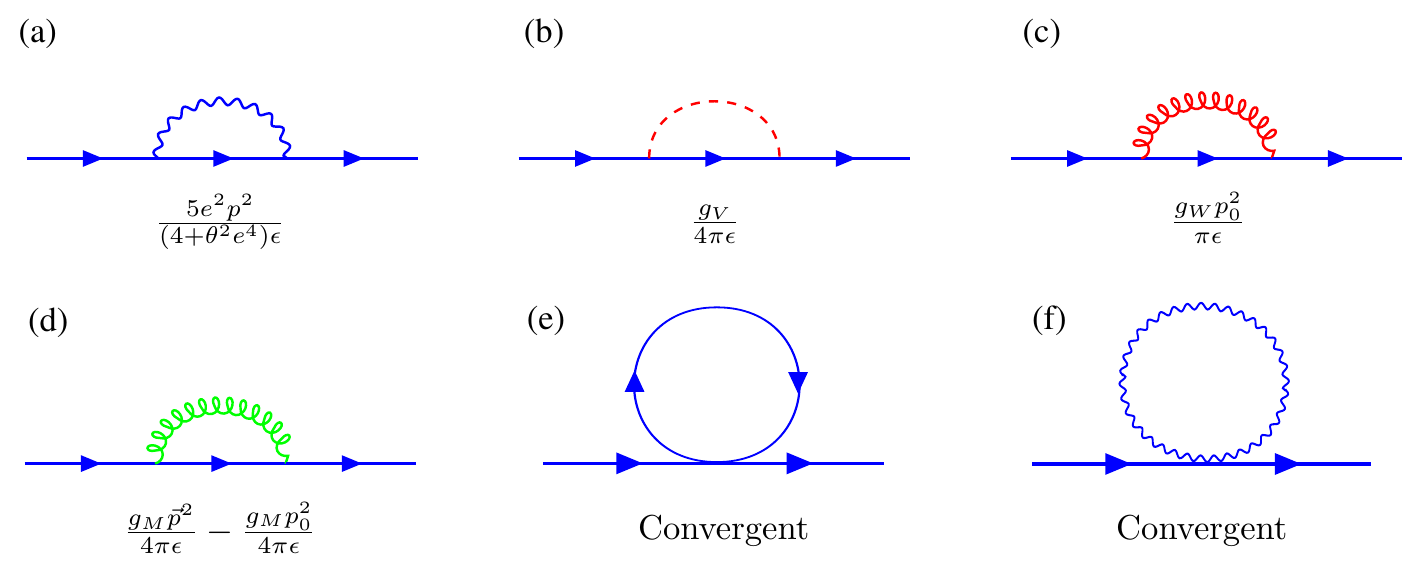}
	\caption{One-loop self-energy diagrams of scalar field with disorder, where we use solid lines for scalar field propagator, sinusoidal lines for optical field propagator, dashed lines for disorder $V(x)$, coiled lines in red for disorder $W(x)$ and coiled lines in green for disorder $M(x)$. If there is a divergent piece in the diagram, it is indicated below the diagram.}
	\label{disself}
\end{figure}

First, we calculate the fermion self-energy diagrams as shown in Fig. (\ref{disself}) (Where we have only list the calculation process of divergence diagram). We use solid line, wavy line, dotted line and circle line remarked by $(0,0)$ and $(i,j)$ represent propagators of scalar field, gauge field, $V(x)$ field, $W(x)$ field and $M_i(x)$ field respectively.

\begin{equation}
	\begin{aligned}
		\mathrm{Fig.\  \ref{disself}(a)}=&\int\frac{d^Dk}{(2\pi)^D}\frac{(2p-k)^\mu(2p-k)^\nu}{(p-k)^2}\Delta^0_{\mu\nu}(k)=\frac{5e^2p^2}{3\pi^2(4+\theta^2e^4)\epsilon}+\mathrm{finite}.
	\end{aligned}
\end{equation}
\begin{equation}
	\begin{aligned}
		\mathrm{Fig.\  \ref{disself}(b)}=&\frac{g_V}{2}\mu^\epsilon\int\frac{d^Dk}{(2\pi)^D}2\pi\delta(k_0)\frac{1}{(p-k)^2}=\frac{g_V}{4\pi\epsilon}+\mathrm{finite}.
	\end{aligned}
\end{equation}	
\begin{equation}
	\begin{aligned}
		\mathrm{Fig.\  \ref{disself}(c)}=&\frac{g_W}{2}\mu^\epsilon\int\frac{d^Dk}{(2\pi)^D}2\pi\delta(k_0)\frac{(2p_0-k_0)^2}{(p-k)^2}=\frac{g_Wp_0^2}{\pi\epsilon}+\mathrm{finite}.
	\end{aligned}
\end{equation}		
\begin{equation}
	\begin{aligned}
		\mathrm{Fig.\  \ref{disself}(d)}=&\frac{g_M}{2}\mu^\epsilon\int\frac{d^Dk}{(2\pi)^D}2\pi\delta(k_0)\delta_{ij}\frac{(2p-k)^i(2p-k)^j}{(p-k)^2}=\frac{g_M\vec{p}^2}{4\pi\epsilon}-\frac{g_Mp_0^2}{4\pi\epsilon}+\mathrm{finite}.
	\end{aligned}
\end{equation}		

These self-energy diagrams determine the counterterms to leading order:
\begin{equation}
	\begin{matrix}
		\delta_1=\frac{5e^2}{3\pi^2(4+\theta^2e^4)\epsilon}+\frac{g_W}{\pi\epsilon}-\frac{g_M}{4\pi\epsilon},	& \delta_2=\frac{5e^2}{3\pi^2(4+\theta^2e^4)\epsilon}+\frac{g_M}{4\pi\epsilon}.
	\end{matrix}
\end{equation}		
We can get the dynamic critical exponent $z=1-\frac{1}{2}\mu \frac{d}{d\mu}\ln\frac{Z_2}{Z_1}=1-\frac{g_M}{4\pi}+\frac{g_W}{2\pi}$.

Diagrams in Fig. \ref{coutV}, Fig. \ref{coutWM}, Fig. \ref{coutU} give the leading order corrections for $g_V$, $g_W$ and $g_M$, $U$, respectively. We listed the calculation process of divergence diagram below:

\begin{figure}
	\includegraphics{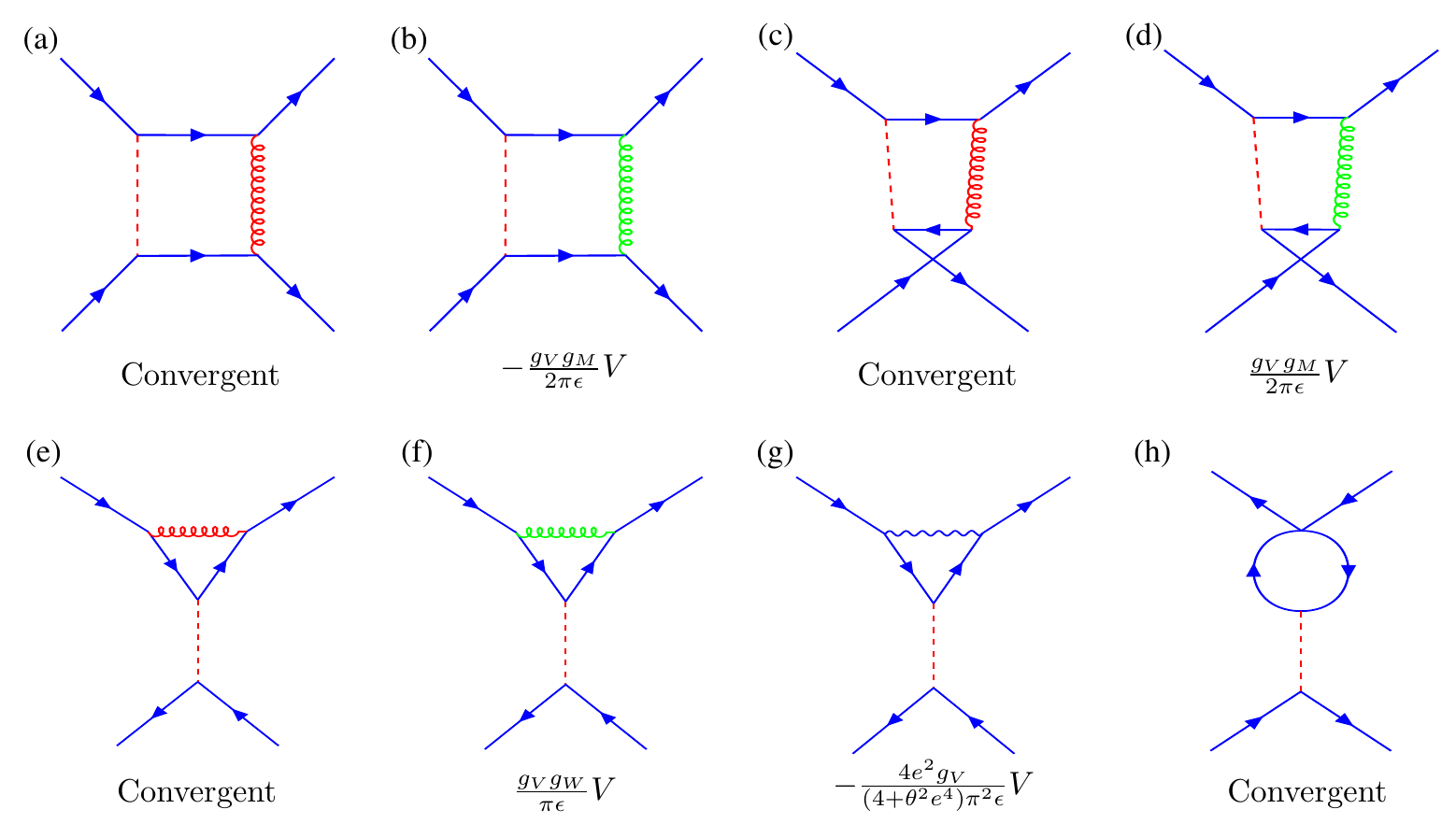}
	\caption{One-loop diagrams contribute to disorder $g_V$, where we use solid lines for scalar field propagator, sinusoidal lines for optical field propagator, dashed lines for disorder $V(x)$, coiled lines in red for disorder $W(x)$ and coiled lines in green for disorder $M(x)$. If there is a divergent piece in the diagram, it is indicated below the diagram. Where $V=2\pi\delta(q^0)$ with $q^0$ the time component of $V(x)$ field momentum.}
	\label{coutV}
\end{figure}

\begin{equation}
	\begin{aligned}
		\mathrm{Fig. \  \ref{coutV}(b)}=&g_Vg_M\mu^{2\epsilon}\int\frac{d^Dq}{(2\pi)^D}2\pi\delta(q_0)2\pi\delta_{ij}\delta(q_0-q_{30}+q_{20})\frac{(q_2+q_3+q)^i(q_2+2q_1-q_3-q)^j}{(q_2+q)^2(q_1-q)^2}\\
		=&-\frac{g_Vg_M2\pi\delta(q_{20}-q_{30})}{2\pi\epsilon}+\mathrm{finite}=-\mathrm{Fig.\   \ref{coutV}(d)},
	\end{aligned}
\end{equation}	
\begin{equation}
	\begin{aligned}
		\mathrm{Fig.\  \ref{coutV}(f)}=&2g_Vg_W\mu^{2\epsilon}\int\frac{d^Dp}{(2\pi)^D}2\pi\delta_{ij}\delta(p_0)2\pi\delta(q_{10}-q_{30})\frac{(2q_1-p)^i(2q_3-p)^j}{(q_1-p)^2(q_3-p)^2}=2\frac{g_Vg_W2\pi\delta(q_{20}-q_{30})}{2\pi\epsilon}+\mathrm{finite},
	\end{aligned}
\end{equation}
\begin{equation}
	\begin{aligned}
		\mathrm{Fig.\  \ref{coutV}(g)}=&4g_V\mu^{\epsilon}\int\frac{d^Dp}{(2\pi)^D}2\pi\delta(q_{10}-q_{30})\frac{(2q_1-p)^\mu(2q_3-p)^\nu}{(q_1-p)^2(q_3-p)^2}\Delta_{\mu\nu}^0(p)=4\frac{e^2g_V2\pi\delta(q_{20}-q_{30})}{(4+\theta^2e^4)\pi^2\epsilon}+\mathrm{finite},
	\end{aligned}
\end{equation}

\begin{figure}[htbp!]
	\includegraphics{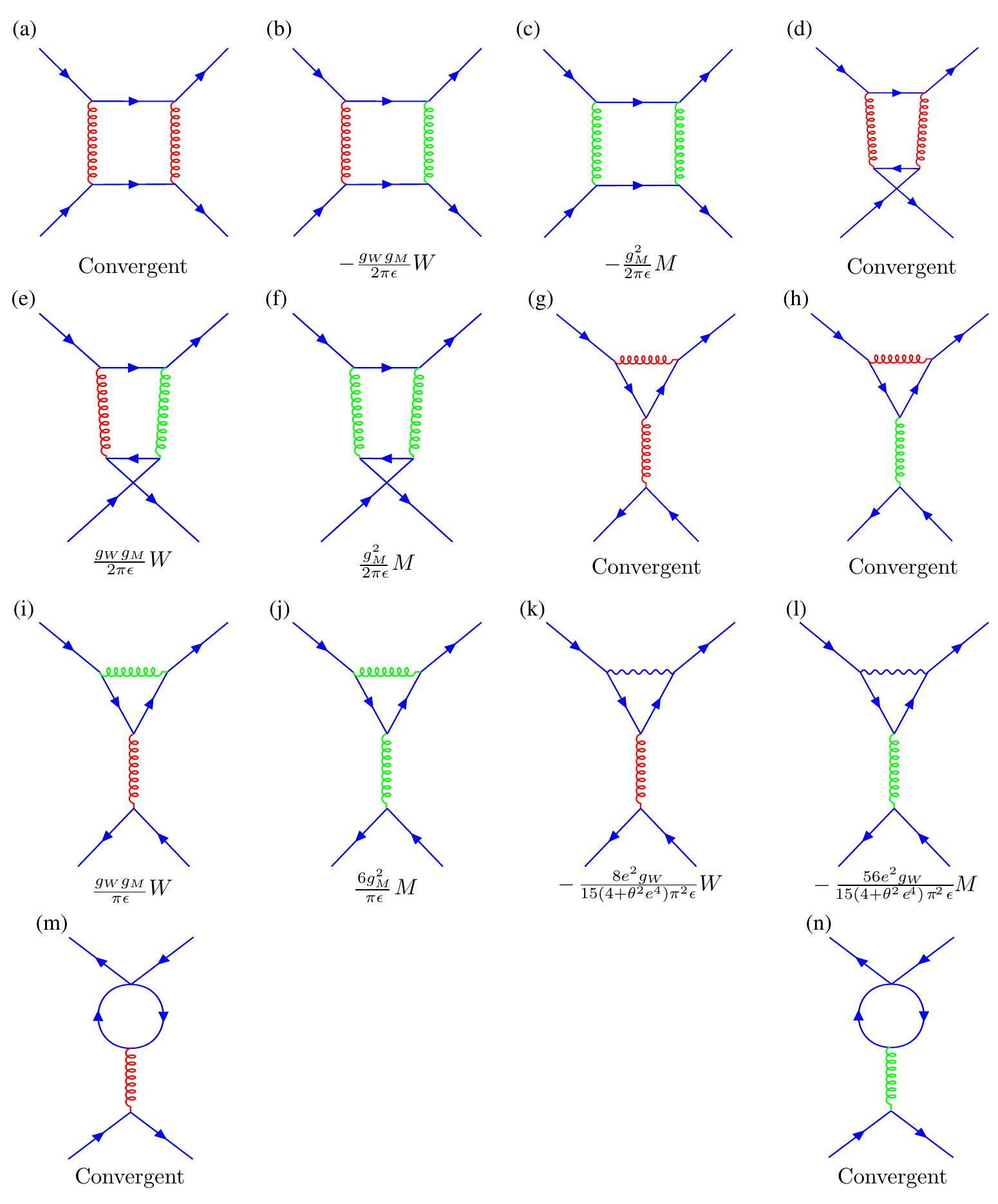}
	\caption{One-loop diagrams contribute to disorder $g_W$ and $g_M$, where we use solid lines for scalar field propagator, sinusoidal lines for optical field propagator, dashed lines for disorder $V(x)$, coiled lines in red for disorder $W(x)$ and coiled lines in green for disorder $M(x)$. If there is a divergent piece in the diagram, it is indicated below the diagram, where $W=8\pi\delta(q_2^0-q_3^0)q_1^0q_2^0$ and $M=2\pi\delta(q_2^0-q_3^0)(q_2^i+q_3^i)(2q_{1i}+q_{2i}-q_{3i})$ with $q_1$, $q_2$ and $q_3$ the momentum of outside line.}
	\label{coutWM}
\end{figure}

\begin{equation}
	\begin{aligned}
		\mathrm{Fig.\  \ref{coutWM}(b)}=&g_Wg_M\mu^{2\epsilon}\int\frac{d^Dq}{(2\pi)^D}2\pi\delta(q_0)2\pi\delta_{ij}\delta(q_0-q_{30}+q_{20})\frac{(q_2+q_3+q)^i(q_2+2q_1-q_3-q)^j(2q_2+q)_0(2q_1-q)_0}{(q_2+q)^2(q_1-q)^2}\\
		=&-\frac{g_Wg_M2\pi\delta(q_{20}-q_{30})4q_{10}q_{20}}{2\pi\epsilon}+\mathrm{finite}=-\mathrm{Fig.\  \ref{coutWM}(e)},
	\end{aligned}
\end{equation}	

\begin{equation}
	\begin{aligned}
		\mathrm{Fig.\  \ref{coutWM}(c)}=&g_M^2\mu^{2\epsilon}\int\frac{d^Dq}{(2\pi)^D}2\pi\delta_{kl}\delta(q_0)2\pi\delta_{ij}\delta(q_0-q_{30}+q_{20})\frac{(q_2+q_3+q)^i(q_2+2q_1-q_3-q)^j(2q_2+q)^k(2q_1-q)^l}{(q_2+q)^2(q_1-q)^2}\\
		=&-\frac{g_M^22\pi\delta(q_{20}-q_{30})(q_2+q_3)_i(2q_1+q_2-q_3)^i}{2\pi\epsilon}+\mathrm{finite}=-\mathrm{Fig.\  \ref{coutWM}(f)},
	\end{aligned}
\end{equation}	

\begin{equation}
	\begin{aligned}
		\mathrm{Fig.\  \ref{coutWM}(i)}=&2g_Mg_W\mu^{2\epsilon}\int\frac{d^Dp}{(2\pi)^D}2\pi\delta_{ij}\delta(p_0)2\pi\delta(q_{10}-q_{30})\frac{(2q_1-p)^i(2q_3-p)^j(q_1+q_3-2p)_0(q_1+2q_2-q_3)_0}{(q_1-p)^2(q_3-p)^2}\\
		&=2\frac{g_Mg_W2\pi\delta(q_{20}-q_{30})4q_{10}q_{20}}{2\pi\epsilon}+\mathrm{finite},
	\end{aligned}
\end{equation}

\begin{equation}
	\begin{aligned}
		\mathrm{Fig.\  \ref{coutWM}(j)}=&12g_M^2\mu^{2\epsilon}\int\frac{d^Dp}{(2\pi)^D}2\pi\delta_{ij}\delta(p_0)2\pi\delta_{kl}\delta(q_{10}-q_{30})\frac{(2q_1-p)^i(2q_3-p)^j(q_1+q_3-2p)_k(q_1+2q_2-q_3)_l}{(q_1-p)^2(q_3-p)^2}\\
		=&12\frac{g_M^22\pi\delta(q_{20}-q_{30})(q_1+q_3)_i(q_1+2q_2-q_3)^i}{2\pi\epsilon}+\mathrm{finite},
	\end{aligned}
\end{equation}

\begin{equation}
	\begin{aligned}
		\mathrm{Fig.\  \ref{coutWM}(k)}=&4g_W\mu^{\epsilon}\int\frac{d^Dp}{(2\pi)^D}2\pi\delta(q_{10}-q_{30})\frac{(2q_1-p)^\mu(2q_3-p)^\nu(q_1+q_3-2p)_0(q_1+2q_2-q_3)_0}{(q_1-p)^2(q_3-p)^2}\Delta_{\mu\nu}^0(p)\\
		=&4\frac{2e^2g_W2\pi\delta(q_{20}-q_{30})4q_{10}q_{20}}{15(4+\theta^2e^4)\pi^2\epsilon}+\mathrm{finite},
	\end{aligned}
\end{equation}

\begin{equation}
	\begin{aligned}
		\mathrm{Fig.\  \ref{coutWM}(l)}=&4g_M\mu^{\epsilon}\int\frac{d^Dp}{(2\pi)^D}2\pi\delta_{ij}\delta(q_{10}-q_{30})\frac{(2q_1-p)^\mu(2q_3-p)^\nu(q_1+q_3-2p)_i(q_1+2q_2-q_3)_j}{(q_1-p)^2(q_3-p)^2}\Delta_{\mu\nu}^0(p)\\
		=&4\frac{14e^2g_M2\pi\delta(q_{20}-q_{30})(q_1+q_3)_i(q_1+2q_2-q_3)^i}{15(4+\theta^2e^4)\pi^2\epsilon}+\mathrm{finite},
	\end{aligned}
\end{equation}

\begin{figure}[htbp!]
	\includegraphics{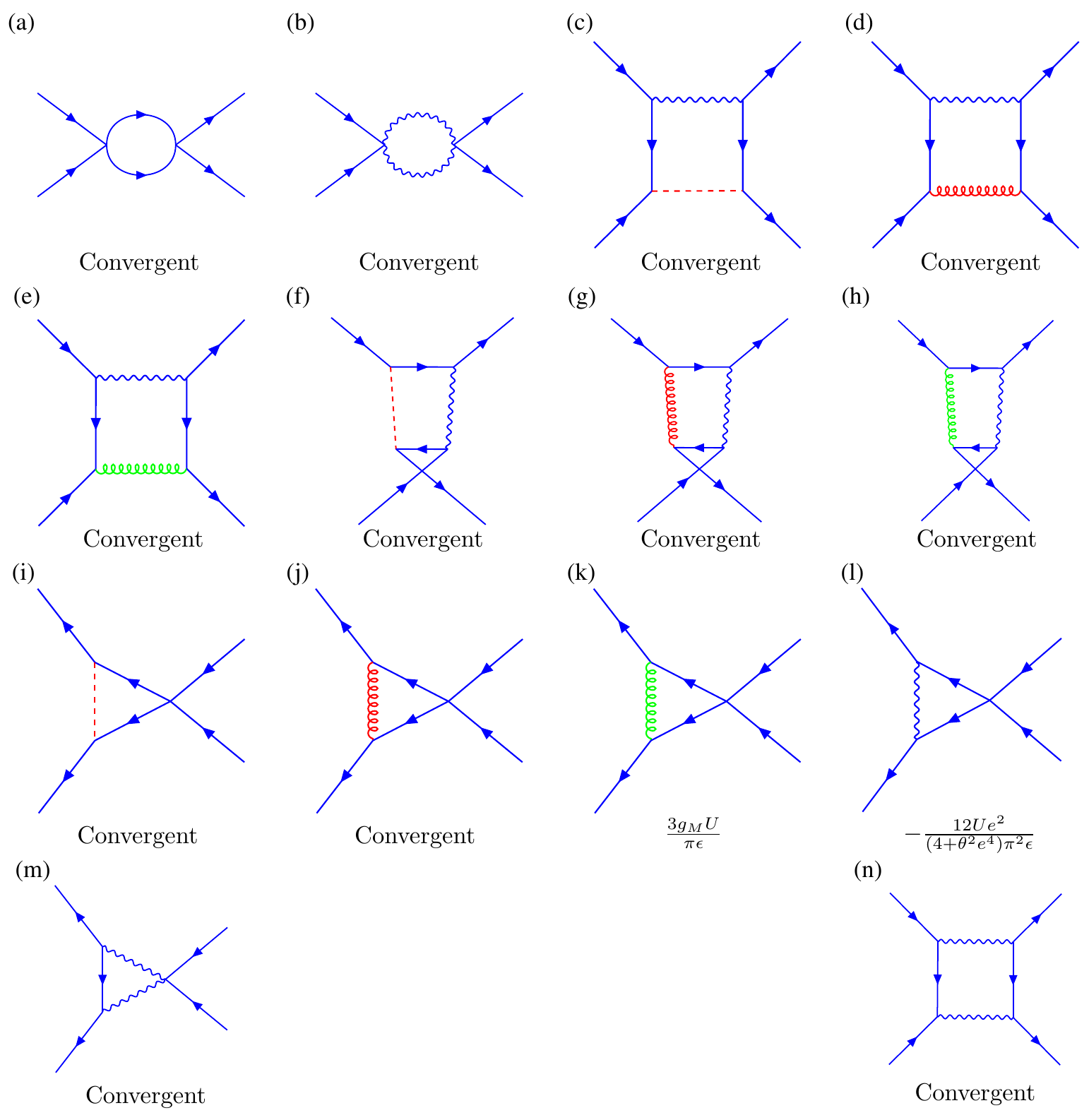}
	\caption{One-loop diagrams contribute to coupling $U$, where we use solid lines for scalar field propagator, sinusoidal lines for optical field propagator, dashed lines for disorder $V(x)$, coiled lines in red for disorder $W(x)$ and coiled lines in green for disorder $M(x)$. If there is a divergent piece in the diagram, it is indicated below the diagram.}
	\label{coutU}
\end{figure}

\begin{equation}
	\begin{aligned}
		\mathrm{Fig.\  \ref{coutU}(k)}=&-6g_MU\mu^{2\epsilon}\int\frac{d^Dq}{(2\pi)^D}2\pi\delta_{ij}\delta(q_{0}+q_{20})\frac{(q_2-q)^i(2p+q-q_2)^j}{q^2(p+q)^2}=6\frac{g_MU}{2\pi\epsilon}+\mathrm{finite},
	\end{aligned}
\end{equation}
\begin{equation}
	\begin{aligned}
		\mathrm{Fig.\  \ref{coutU}(l)}=&-12U\mu^{\epsilon}\int\frac{d^Dq}{(2\pi)^D}\frac{(q_2-q)^\mu(2p+q-q_2)^\nu}{q^2(p+q)^2}\Delta_{\mu\nu}^0(q_2+q)=12\frac{Ue^2}{(4+\theta^2e^4)\pi^2\epsilon}+\mathrm{finite},
	\end{aligned}
\end{equation}

We can get the counterterms from above results:
\begin{equation}
	\begin{aligned}
		\delta_V=&-\frac{2g_Ve^2}{(4+\theta^2 e^4)\pi^2\epsilon}-\frac{g_Vg_W}{2\pi\epsilon},\ \ \delta_W=-\frac{4g_We^2}{15(4+\theta^2 e^4)\pi^2\epsilon}-\frac{g_Mg_W}{2\pi\epsilon},\\
		\delta_M=&-\frac{28g_Me^2}{15(4+\theta^2 e^4)\pi^2\epsilon}-\frac{3g_M^2}{\pi\epsilon},\ \ \delta_U=\frac{6Ue^2}{(4+\theta^2e^4)\pi^2\epsilon}+\frac{3g_MU}{2\pi\epsilon}.
	\end{aligned}
\end{equation}
The beta function of $g_V$, $g_W$, $g_M$ and $U$ can be obtained from the fllowing relation:
\begin{equation}
	g_{V,B}=\mu^{2+\epsilon}Z_2^{-2}(g_V+\delta_V),\ \ g_{W,B}=\mu^\epsilon Z_2^{-1}Z_1^{-1}(g_W+\delta_W),\ \ g_{M,B}=\mu^{\epsilon}Z_2^{-2}(g_M+\delta_M),\ \   U_B=\mu^\epsilon Z_2^{-2}(U+\delta_U).
\end{equation}

Simple scaling analysis indicates that the critical behaviors are primarily determined by $g_W$, $g_M$, and $c$ due to their zero scaling dimension. The non-zero scaling dimensions of $U$ and $g_V$ indicate the need for additional terms to offset the redundancies in the counterterms. The only possible choice is for $m$ to appear in the denominator, but we do not have a mass term. So $U$ and $g_V$ do not contribute to higher-order terms that appear in the beta functions. Another possible gauge-invariant disorder operator is $|D_\mu\phi|$ which could significantly affect the disorder critical behavior. However, we can fix the kinetic term and attribute the effect of disorder to other terms, which leads to the same result as above.

\section{Vortices and duality}

\subsection{One way to derive the effective action describing vortex worldlines from plane}
To discuss the properties of vortices, we first change the complex scalar field to polar coordinates $\phi=\rho e^{i\theta}$ with a superfluid density $|\rho|^2$ and phase $\theta$ and get the effective action like:
\begin{equation}
	\mathcal{L}_{eff}=|\rho|^2(\partial_\mu\theta-A_\mu)^2+r|\rho|^2 +\frac{U}{2}|\rho|^4+\frac{1}{4e^2}F^{\mu\nu}\frac{2}{\sqrt{\Box}}F_{\mu\nu}+i\frac{\theta}{2}\epsilon^{\mu\nu\rho}A_\mu\partial_\nu A_\rho.
\end{equation}
Using the Hubbard-Stratonovich transformation by introducing the auxiliary fields $h_\mu$. Then $\theta$ is decomposed into phase part (regular part) $\chi$ and the vortex gauge field (singular parts) $\nu_\mu$, $\partial_\mu\theta=\partial_\mu \chi +2\pi\nu_\mu$ (We note that the decompositions of the phase of the scale field is determined by the spatial phase shift of vortex. It is $2\pi$ in our case. The spatial phase shift in pair density wave order is $ \pi $  as found in Ref. \cite{liu2023pair}, the decomposition can be written as  $\partial_\mu \theta=\partial_\mu \chi +\pi \nu_\mu$.), we get:
\begin{equation}
	\mathcal{L}_{eff}'=\frac{1}{4|\rho|^2}h_\mu^2-ih_\mu(\partial_\mu \chi+2\pi  \nu_\mu-A_\mu)+r|\rho|^2 +\frac{U}{2}|\rho|^4+\frac{1}{4e^2}F^{\mu\nu}\frac{2}{\sqrt{\Box}}F_{\mu\nu}+i\frac{\theta}{2}\epsilon^{\mu\nu\rho}A_\mu\partial_\nu A_\rho.
\end{equation}
After integrating out the gauge field $A_\mu$, we can get the effective action:
\begin{equation}
	\mathcal{S}_{eff}'=\int d^3x [\frac{1}{4|\rho|^2}h_\mu^2+r|\rho|^2 +\frac{U}{2}|\rho|^4-ih_\mu(\partial_\mu \chi+2\pi  \nu_\mu)+\frac{1}{2}\int d^3x' h_\mu(x)\Delta_{\mu\nu}^0(x-x')h_\nu(x')].
	\label{sa}
\end{equation}
Then integrating out the regular part $\chi$, we can get the constraint $\partial_\mu h_\mu=0$, there are two solitions satisfied the unitarity \cite{WOS:000488071200096}. The first one is $h_\mu=\epsilon_{\mu\nu\lambda}\partial_\nu a_\lambda/2\pi$. Insert it into the action (\ref{sa}):
\begin{equation}
	\begin{aligned}	
		\mathcal{L}_{eff}=&\frac{1}{16\pi^2|\rho|^2}(\epsilon_{\mu\nu\lambda}\partial_\mu a_\lambda)^2+r|\rho|^2 +\frac{U}{2}|\rho|^4-i a_\lambda J_\lambda^a+\frac{e^2}{4\pi^2(4+\theta^2e^4)}\frac{(\epsilon_{\mu\nu\lambda}\partial_\mu a_\lambda)^2}{\sqrt{\Box}}-i\frac{\theta e^4}{8\pi^2(4+\theta^2e^4)}\epsilon_{\mu\nu\lambda}a_\mu\partial_\nu a_\lambda,
		\label{eq}
	\end{aligned}
\end{equation}
where $J_\mu^a=\epsilon_{\mu\nu\lambda}\partial_\nu \nu_\lambda$ is the vortex current. $a_\mu$ can be understood as a $U(1)$ gauge field because $a_\mu\rightarrow a_\mu+\partial_\mu \lambda$ does not change $h_\mu$. At the same time, the Lagrangian (\ref{eq}) is gauge invariant with topologically trivial gauge transformation.  

The other solution is $h_\mu=\epsilon_{\mu\nu\lambda}\partial_\nu[a_\lambda/(\Box)^{1/4}]/2\pi$, insert it into the action (\ref{sa}):
\begin{equation}
	\begin{aligned}	
		\mathcal{L}_{eff}=&\frac{1}{16\pi^2|\rho|^2}\frac{(\epsilon_{\mu\nu\lambda}\partial_\mu a_\lambda)^2}{\sqrt{\Box}}+r|\rho|^2 +\frac{U}{2}|\rho|^4-i \frac{a_\lambda J_\lambda^a}{(\Box)^{\frac{1}{4}}}+\frac{e^2}{4\pi^2(4+\theta^2e^4)}\frac{(\epsilon_{\mu\nu\lambda}\partial_\mu a_\lambda)^2}{\Box}-i\frac{\theta e^4}{8\pi^2(4+\theta^2e^4)}\frac{\epsilon_{\mu\nu\lambda}a_\mu\partial_\nu a_\lambda}{\sqrt{\Box}}.
	\end{aligned}
	\label{eq2}
\end{equation}
We will not accepet the second solution because the non-local Chern-Simon term is not quantized. We note that the scalar dimension of the dual gauge field $a_u$ in Lagrangian (\ref{eq}) is the same as that of the original background gauge field $A_\mu$, while in Lagrangian (\ref{eq2}), it is not.

\subsection{Another way to derive the effective action describing vortex worldlines from bulk}

We will get the same results from the electric–magnetic duality \cite{WOS:000468428800003,seiberg2016duality}. Consider the dynamical $U(1)$ gauge ﬁeld $A_\mu$ on $\mathbb{R}^3\times \mathbb{R}_+$. The (2+1)D system we consider to be located at $y\equiv x^3=0$, and we denote it by M. We denote $x^\mu$, $\mu$=0,1,2,3, as coordinates on the whole space $\mathbb{R}^3\times \mathbb{R}_+$ and $x^i$, $i$=0,1,2, as coordinates on the planar system M.  The Maxwell term of the whole space can be written as $F^{\mu\nu}F_{\mu\nu}/4e^2$ and the CS term, $\theta \epsilon^{ijk}A_i\partial_j A_k /2 $, has the form $\theta \epsilon^{\mu\nu\rho \theta}\partial_\mu A_\nu \partial_\rho A_\theta/4=\theta \epsilon_{\mu\nu\rho \theta}F^{\mu\nu}F^{\rho \theta}/16$. We assume the gynamical gauge field $A_\mu$ limited at the boundary M of the half-apcetime and coupling to some degrees of freedom that propagate on M, which can be seen as background field. The effective Lagrangian can be written as:
\begin{equation}
	\begin{aligned}
		&\int_M \mathcal{L}(\phi,A_i)+\int_{y\ge 0}dy d^3x(\frac{1}{4e^2}F^{\mu\nu}F_{\mu\nu}+\frac{i\theta}{16}\epsilon_{\mu\nu\rho \theta}F^{\mu\nu}F^{\rho \theta})\\
		=&\int_M \mathcal{L}(\phi,A_i)-\frac{i}{8\pi}\int_{y\ge 0}dy d^3x(\tau F^-_{\mu\nu}F^{\mu\nu-}-\bar{\tau}F^+_{\mu\nu}F^{\mu\nu+})+\frac{i}{16\pi}\int_{y\ge 0}dy d^3x \epsilon_{\mu\nu}^{\ \ \rho \theta}B^{\mu\nu}(F_{\rho \theta}-\partial_\rho A_\theta+\partial_\theta A_\rho),
	\end{aligned}
\end{equation}
where $\tau=\frac{i2\pi}{e^2}+\pi\theta$ and $F_{\mu\nu}^\pm=(F_{\mu\nu}\pm \epsilon_{\mu\nu}^{\ \ \rho\theta}F_{\rho\theta}/2)/2$ is the self-dual/anti-self-dual field strength tensor. The last term gives the constraint setting $F_{\mu\nu}=\partial_\mu A_\nu-\partial_\nu A_\mu$ after path integraling over the $B_{\mu\nu}$. At this time, we can treat $F_{\mu\nu}$ as an independent variable of $A_\mu$ and integrating over it:
\begin{equation}
	\int_M \mathcal{L}(\phi,A_i)-\frac{i}{8\pi}\int_{y\ge 0}dy d^3x(\tau' B^-_{\mu\nu}B^{\mu\nu-}-\bar{\tau}'B^+_{\mu\nu}B^{\mu\nu+})-\frac{i}{8\pi}\int_{y\ge 0}dy d^3x \epsilon_{\mu\nu}^{\ \ \rho \theta}B^{\mu\nu}\partial_\rho A_\theta,
\end{equation}
where $\tau'=-1/\tau$. $B_{\mu\nu}=\partial_\mu B_\nu-\partial_\nu B_\mu$ can be viewed as new field strength tensor with a $U(1)$ gauge field $B_\mu$. Restrict redundant items, $\frac{i}{8\pi}\int_{y\ge 0}dy d^3x \epsilon_{\mu\nu}^{\ \ \rho \theta}B^{\mu\nu}\partial_\rho A_\theta$, to planes, we get $\frac{i}{2\pi}\int_M \epsilon^{ijk }A_i\partial_j B_k$. 

The dual Lagrangian with dual gauge field $B_\mu$ is:
\begin{equation}
	\begin{aligned}
		\mathcal{L}_{dual}=&\frac{e^2}{4\pi^2(4+\theta^2e^4)}B^{ij}\frac{2}{\sqrt{\Box}}B_{ij}+|(\partial_i -iA_i)\phi|^2+r|\phi|^2 +\frac{U}{2}|\phi|^4-i\frac{\theta e^4}{2\pi^2(4+\theta^2e^4)}\epsilon^{ijk}B_i\partial_j B_k-\frac{i}{2\pi}\epsilon^{ijk }A_i\partial_j B_k.
	\end{aligned}
	\label{ph}
\end{equation} 

Similar process as before, we change the scalar field $\phi$ into polar coordinates and integrate out the dunamical gauge field $A_i$. We get:
\begin{equation}
	\mathcal{L}_{eff}=\frac{1}{4\pi^2|\rho|^2}(\epsilon^{ijk}\partial_i B_j)^2+r|\rho|^2 +\frac{U}{2}|\rho|^4-i2 B^i J_i^a+\frac{e^2}{\pi^2(4+\theta^2e^4)}\frac{(\epsilon^{ijk}\partial_i B_j)^2}{\sqrt{\Box}}-i\frac{\theta e^4}{2\pi^2(4+\theta^2e^4)}\epsilon^{ijk}B_i\partial_j B_k.
	\label{emd}
\end{equation}
Rescale $B_\mu\rightarrow B_\mu/2$ we get the same result as shown in Lagrangian (\ref{eq}).

Because gauge symmetry is broken, the Lagrangian in (\ref{eq}) or (\ref{emd}) describes an insulator. The superfluidity with pseudoparticle paramagnetic current $J_\mu^b=2|\rho|^2\partial_\mu \chi$ is described by \cite{burgess2001particle}:
\begin{equation}
	\mathcal{L'}_{eff}=\frac{(J_\mu^{b})^2}{4|\rho|^2}-J_\mu^b A^\mu+|\rho|^2A_\mu^2++r|\rho|^2 +\frac{U}{2}|\rho|^4+\frac{1}{4e^2}F^{\mu\nu}\frac{2}{\sqrt{\Box}}F_{\mu\nu}+i\frac{\theta}{2}\epsilon^{\mu\nu\rho}A_\mu\partial_\nu A_\rho.
\end{equation}

The propagator of dual gauge field $a_\lambda $ in Lagrangian (\ref{eq}) is:
\begin{equation}
	\begin{aligned}
		\tilde{\Delta}_{\alpha \beta}(p^2)/4\pi^2&=\frac{\frac{p^2}{2|\rho|^2}+\frac{2e^2}{4+\theta^2e^4}\sqrt{p^2}}{(\frac{\theta e^4}{4+\theta^2e^4})^2p^2+(\frac{p^2}{2|\rho|^2}+\frac{2e^2}{4+\theta^2e^4}\sqrt{p^2})^2}(\delta_{\alpha \beta}-p_\alpha p_\beta/p^2)-\frac{\frac{\theta e^4}{4+\theta^2e^4}}{(\frac{\theta e^4}{4+\theta^2e^4})^2p^2+(\frac{p^2}{2|\rho|^2}+\frac{2e^2}{4+\theta^2e^4}\sqrt{p^2})^2}\epsilon_{\alpha \beta \nu}p_\nu.
	\end{aligned}
\end{equation}
After integrating over the dual gauge field $a_\mu$ in the limit of $|\rho|^2\gg p^2$, we obtain the effection action:
\begin{equation}
	\begin{aligned}	
		S_{eff}=&\int d^3x d^3x'
		[-\frac{4\pi^2}{e^2}\frac{J_\mu^a(x) J^{\mu a}(x')}{2\pi^2|x-x'|^2}+i2\pi^2 \theta\frac{\epsilon^{ \mu\alpha\nu} J_\mu^a(x)(x-x')_\alpha 		J_\nu^a(x')}{4\pi|x-x'|^3}].
		\label{eact}
	\end{aligned}
\end{equation}
Substitute the vortex current $J_\mu^a(x)=\epsilon_{\mu\nu\lambda}\partial^\nu v^\lambda(x)=\sum_cn_c\oint_{L_c}dy^{c}_\mu\delta^3(x-y^{c})$ into it, we obtain:
\begin{equation}
	\begin{aligned}
		S_{eff}&=-\frac{2}{e^2}\sum_{c,c'}n_cn_{c'}\oint_{L_c}dy_\mu^c\oint_{L_{c'}}dz^{\mu c'}\frac{1}{|y^c-z^{c'}|^2}+\frac{i\pi\theta }{2}\sum_{c,c'}n_cn_{c'}\oint_{L_c}dy_\mu^c\oint_{L_{c'}}dz_\nu^{ c'}\epsilon^{\mu\nu\alpha}\frac{y_\alpha^c-z_\alpha^{c'}}{|y^c-z^{c'}|^3}
	\end{aligned}
\end{equation}
The definition of the linking number $G$, the writhe $W$ and the twist $T$ of two curves $L_c$ and $L_{c'}$ can be found in \cite{turker2020bosonization}. The second term at right hand side can character $2\pi^2\theta$ times the linking number of two different vortex loops or the writhe of the same loop. It captures the phase obtained after one vortex circles another. Such the effective self-statistical angle of unit vortex is $\pi n$, satisfied Fermi statistics at $n=1$.

\subsection{The dual Lagrangian of the original one}

Conservation of current $J_\mu$, $\partial^\mu J_\mu=0$, indicates we can add the action, $i\partial^\mu J_\mu \xi$, as a constraint setting. Then introducing a convergence factor \cite{peskin1978mandelstam}, $tJ_\mu^2/2$, and integrating over $J_\mu$, we obtain:
\begin{equation}
	\begin{aligned}	
		\mathcal{\tilde{L}}=\frac{e^2}{4\pi^2(4+\theta^2e^4)}\frac{(\epsilon^{\mu\nu\lambda}\partial_\mu a_\lambda)^2}{\sqrt{\Box}}-i\frac{\theta e^4}{8\pi^2(4+\theta^2e^4)}\epsilon^{\mu\nu\lambda}a_\mu\partial_\nu a_\lambda+\frac{1}{2t}(\partial_\mu \xi-a_\mu)^2+\frac{1}{16\pi^2|\rho|^2}(\epsilon^{\mu\nu\lambda}\partial_\mu a_\lambda)^2+r|\rho|^2 +\frac{U}{2}|\rho|^4.
		\label{eff2}
	\end{aligned}
\end{equation}
It has the same formal structure at long wavelengths:
\begin{equation}
	\begin{aligned}
		\mathcal{\tilde{L}}_{dual}=&\frac{1}{4\tilde{e}^2}\tilde{F}^{\mu\nu}\frac{2}{\sqrt{\Box}}\tilde{F}_{\mu\nu}+|(\partial_\mu -i\tilde{A}_\mu)\tilde{\phi}|^2+\tilde{r}|\tilde{\phi}|^2+\frac{\tilde{U}}{2}|\tilde{\phi}|^4+i\frac{\tilde{\theta}}{2}\epsilon^{\mu\nu\rho}\tilde{A}_\mu\partial_\nu \tilde{A}_\rho,
	\end{aligned}
	\label{lad}
\end{equation}
where $\tilde{\theta}=-\theta e^4/(4\pi^2 (4+\theta^2e^4))$ and $\tilde{e}^2=\pi^2(4+\theta^2e^4)/e^2$. 

To better understand the infrared duality between the original Lagrangian (1) and the Lagrangian (\ref{lad}), we rewrite it as:
\begin{equation}
	\begin{aligned}
		\mathcal{\tilde{L}}_{dual}=|(\partial_\mu -ia_\mu)\tilde{\phi}|^2+\tilde{r}|\tilde{\phi}|^2+\frac{\tilde{U}}{2}|\tilde{\phi}|^4-\frac{i}{2\pi}\epsilon^{\mu\nu\gamma} a_\mu\partial_\nu \tilde{A}_\gamma,
	\end{aligned}
	\label{ladd}
\end{equation}
with the dynamical gauge field $a_\mu$ and the background gauge field $\tilde{A}_\mu$. In Lagrangian (\ref{lad}), we constrain the background gauge field $\tilde{A}_\mu$ on the plane (surface), while we make it lives in the full space in Lagrangian (\ref{ladd}).

The positive $r$ in Lagrangian (1) makes $\phi$ gapped and the global $U(1)_A$ symmetry is unbroken. If we want get the same physics in Lagrangian (\ref{ladd}), we need negative $\tilde{r}$ to generate gapped vortrex excitations (Higgs mechanism). Moreover, these excitations carry unit charge due to the unbroken $U(1)_{\tilde{A}}$ symmetry. The negative $r$ in Lagrangian (1) brokes the global $U(1)_A$ symmetry and induces a Goldstone boson. It is dual to positice $\tilde{r}$, the gauge field $a_\mu$ is massless and can be seen as Goldstone boson of broken $U(1)_{\tilde{A}}$ symmetry.  

The previous analysis in the main text indicates that our model has a second-order phase transition. We can define a conformal field theory at the quantum critical point. At temperature $T=0$, the retarded correlator $C_{\mu \nu}$ of the conserved currents $j_\mu$  is
\begin{equation}
	C_{\mu \nu}\left( p \right) =K\sqrt{p^2}\left( \eta _{\mu \nu}-\frac{p_{\mu}p_{\nu}}{p^2} \right) +H\epsilon _{\mu \nu \lambda}p^{\lambda},
\end{equation}
where $K$, $H$  are two real constants and this form preserves Lorentz and scale invariance. We can also have a dual correlator $\widetilde{C_{\mu \nu}}$  with parameters $\tilde{K}$  and  $\tilde{H}$ in dual Lagrangian (\ref{ladd}). The exact relation of these parameters at $T=0$ is:
\begin{equation}
	\left( H+iK \right) \left( \tilde{H}+i\tilde{K} \right) =-\frac{1}{4\pi ^2}
\end{equation}
at the limit of $e\rightarrow \infty$ and $\tilde{e}\rightarrow \infty$. The self-duality requires $H=\tilde{H}$ and $K=\tilde{K}$. We have  $H=0$ and $K=1/2\pi$. It determines the flow correlation function near the critical point. 

This relation can be attended to the case of $T>0$  where the Lorentz invariance is replaced by spatial rotational invariance. The retarded correlator
\begin{equation}
	C_{\mu \nu}\left( p \right) =\sqrt{p^2}\left[ K^T\left( \omega ,k \right) P_{\mu \nu}^{T}+K^L\left( \omega ,k \right) P_{\mu \nu}^{L}+H\left( \omega ,k \right) \epsilon _{\mu \nu \lambda}p^{\lambda} \right] ,
\end{equation}
where the $P_{\mu \nu}^{T}$ and $P_{\mu \nu}^{L}$ are orthogonal projectors of  $\eta _{\mu \nu}-p_{\mu}p{{_{\nu}}/{p^2}}$ in the spatial and residual parts, respectively. The self-duality reduces to the condition:
\begin{equation}
	K^T\left( \omega ,k \right) K^L\left( \omega ,k \right) =\frac{1}{4\pi ^2},\ \ \ H\left( \omega ,k \right) =0.
\end{equation}
Other constraints are needed to obtain the correlation function for non zero temperature.

\end{document}